\documentclass[eqsecnum,tightenlines,superscriptaddress,nofootinbib,10pt,twocolumn,raggedbottom]{revtex4-2}

\usepackage[english]{babel}

\usepackage[letterpaper,top=2cm,bottom=2cm,left=3cm,right=3cm,marginparwidth=1.75cm]{geometry}

\usepackage[colorlinks=true, allcolors=blue]{hyperref}
\usepackage{algpseudocode}
\usepackage{amsmath,amsfonts,amssymb,amsthm}
\usepackage{graphicx}
\usepackage{adjustbox}
\usepackage{microtype}
\usepackage{needspace}
\usepackage{float}
\usepackage{subcaption}
\usepackage{dsfont}
\usepackage{color}
\usepackage[normalem]{ulem}
\usepackage{multirow}  
\usepackage{empheq}
\usepackage{algorithm}
\usepackage{svg}
\usepackage{booktabs}
\usepackage{braket}
\usepackage{cleveref}
\usepackage{physics}
\newcommand{\SU}{\mathrm{SU}}

\newcommand{\referencegraphic}[2][\refscale]{%
  \begin{adjustbox}{max width=\linewidth}%
    \includegraphics[scale=#1]{#2}%
  \end{adjustbox}%
}
\newcommand{\refpair}[6]{%
  \par\noindent
  \begin{minipage}[b]{0.49\textwidth}\centering\referencegraphic{#1}\end{minipage}\hfill
  \begin{minipage}[b]{0.49\textwidth}\centering\referencegraphic{#4}\end{minipage}%
  \par\nobreak\noindent
  \begin{minipage}[t]{0.49\textwidth}\captionof{figure}{#2}\label{#3}\end{minipage}\hfill
  \begin{minipage}[t]{0.49\textwidth}\captionof{figure}{#5}\label{#6}\end{minipage}%
  \par\vspace{14pt}%
}
\newcommand{\refsingle}[4][\refscale]{%
  \par\noindent
  \begin{minipage}[b]{\textwidth}\centering\referencegraphic[#1]{#2}\end{minipage}%
  \par\nobreak\noindent
  \begin{minipage}[t]{\textwidth}\captionof{figure}{#3}\label{#4}\end{minipage}%
  \par\vspace{14pt}%
}


\graphicspath{ {./Plots/}}

\makeatletter
\def\fnum@figure{FIG~\thefigure}
\makeatother

\makeatletter
\def\fnum@table{TABLE~\thetable}
\makeatother

\crefname{equation}{Eq.}{Eqs.}
\crefname{section}{Sec.}{Secs.}
\crefname{figure}{Fig.}{Figs.}
\crefname{appendix}{Appendix}{Appendices}
\crefname{algorithm}{Algorithm}{Algorithms}
\crefname{table}{Table}{Tables}

\crefformat{table}{(Table~#2#1#3)}

\begin{document}


\title{Probing the semiclassical regime of spinfoams: a numerical study of SU(2) BF theory}

\author{Pietro Dona}
\email{pietro.dona@cpt.univ-mrs.fr}
\affiliation{Aix-Marseille Univ, Universit\'e de Toulon, CNRS, CPT, Marseille, France}

\author{Athanasios Kogios}%
 \email{akogios@perimeterinstitute.ca}
\affiliation{Perimeter Institute, 31 Caroline Street North, Waterloo, Ontario, N2L 2Y5, Canada}
\affiliation{Department of Physics and  Astronomy, University of Waterloo, 200 University Avenue West, Waterloo, Ontario, N2L 3G1, Canada}


\begin{abstract}
Spinfoam amplitudes are intended to provide a path-integral description of transitions between loop quantum gravity states. We use numerical methods to explore the regime of classical behavior in four-dimensional $\SU(2)$ BF theory. We consider symmetric triangulations made of 4--simplices glued around a common triangle. For each configuration, we fix a coherent boundary state. This state describes the boundary of a classical Euclidean Regge geometry with a prescribed deficit angle. We compute the amplitudes with graphical recoupling methods and numerical vertex evaluations performed with \texttt{sl2cfoam-next}. We then compare the amplitude-weighted expectation values of bulk spins and internal intertwiners with their classical references. We also compute their relative dispersions. We define as classical the regime where the expectation values match the geometric ones and the dispersions are small. We study how the agreement with the classical reference depends on the deficit angle, the spin scale, and the size of the triangulation. This dependence highlights the interplay between intertwiner dynamics and phase cancellations in extended spinfoam amplitudes.
\end{abstract}

\maketitle


\twocolumngrid

\clearpage
\section{Introduction}
\label{sec:ch5:introduction}

Spinfoam models define covariant transition amplitudes for loop quantum gravity as sums over discrete quantum geometries \cite{Perez:2012wv,Rovelli:2014ssa}. Their semiclassical limit is therefore a central consistency question: when all spins are large, the amplitudes should select Regge geometries \cite{Regge:1961px,Han:2023cen} and reproduce the classical gravitational dynamics in the appropriate regime. This expectation is on firm ground for a single four-simplex. In particular, the EPRL-FK vertex \cite{Engle:2007uq,Engle:2007wy,Freidel:2007py} has a large-spin asymptotic expansion governed by the Regge action \cite{Barrett:2009gg,Barrett:2009mw}, and its numerical behavior has been studied in detail for individual vertices \cite{Dona:2019dkf}. The harder question is what happens for an extended simplicial complex. Once several vertices are glued together, internal spins and intertwiners are summed over, and the resulting interference can shift, broaden, or suppress the configurations that dominate the amplitude. Here we explore this mechanism numerically, testing whether a controlled multi-vertex amplitude still selects the expected bulk geometry.

We investigate this question in four-dimensional $\SU(2)$ BF theory. We choose this conservative setting intentionally to keep the numerical calculations as simple as possible. BF theory is topological: its discrete amplitudes impose flatness and do not describe local gravitational degrees of freedom \cite{Ponzano:1969tw,Ooguri:1992eb}. For that reason, it is not a substitute for the gravitational EPRL-FK model. It is, however, a useful intermediate problem for understanding the mathematical and numerical tools that we will need for the physical problem. The $\SU(2)$ BF amplitude already contains the ingredients that make extended spinfoam calculations delicate: oscillatory large-spin behavior, nontrivial recoupling theory, finite but rapidly growing internal sums, and cancellations between contributions with different phases. Studying this model first lets us separate difficulties caused by the composition of $\SU(2)$ amplitudes from those introduced later by the simplicity constraints of the gravitational theory.

The calculation is motivated by double-scaling approaches to the semiclassical and continuum regime of spinfoam gravity, which combine large boundary spins with an increasing number of simplices \cite{Han:2013hga,Han:2017cwc,Han:2021kmi}. The interplay between these two regimes has also been probed numerically within effective spin foam models \cite{Asante:2020txa,Asante:2021pxr,Asante:2020sfl}. Here we test for semiclassical behavior at finite, moderately large boundary spins, studying its dependence on the number of four-simplices in a family of triangulations $\Delta_n$. Each triangulation is formed by gluing four-simplices around a common triangle. The spin on this shared triangle, denoted $j_{\text{b}}$, is the main observable. The setup is deliberately chosen so that there is a clear classical target. We assign coherent boundary data peaked on the boundary geometry of a triangulation with prescribed deficit angle $\delta$, and test the relation between the classical area of the shared triangle and the expectation value of the corresponding quantum operator. This gives a direct test for the quantum amplitude: if the amplitude is selecting the intended geometry and we are in the semiclassical regime, the expectation value $\expval{j_{\text{b}}}$ (in the path-integral sense) should follow the classical spin associated with that shared triangle.

The $\SU(2)$ BF and EPRL-FK models share the same boundary Hilbert space: both are built on the kinematical Hilbert space of loop quantum gravity, spanned by $\SU(2)$ spin networks \cite{Perez:2012wv}, with coherent intertwiners giving these states a geometric interpretation in terms of glued discrete polyhedra \cite{Livine:2007vk}. The two models can therefore be understood as different bulk dynamics acting on the same boundary states, with the EPRL-FK simplicity constraints distinguishing the gravitational theory from the topological BF theory studied here.

A further issue is the intertwiner's role. In the recoupling basis used in the computation, the internal tetrahedra carry intertwiner labels in addition to the bulk spin. These labels encode shape degrees of freedom, and they are not fixed automatically by fixing the boundary spins and coherent boundary geometry. This is one of the paper's main points. We first evaluate the unrestricted amplitude, allowing the intertwiner to range over all admissible values. We then introduce a projected amplitude in which the internal intertwiner is restricted to a finite window around its classical value. Because these results are obtained at large but finite spins, read them as evidence for the onset of semiclassical behavior rather than as a strict asymptotic proof. The calculation is best understood as diagnostic: it points to which variables matter and how they respond as the number of four-simplices increases.

The rest of the paper is organized as follows. \cref{sec:ch5:model} defines the $\SU(2)$ BF amplitude used in the computation. \cref{sec:ch5:config} introduces the triangulations, boundary data, and observables. \cref{sec:ch5:numerics} describes the numerical implementation and the intertwiner projector. \cref{sec:ch5:results} presents the bulk spin and intertwiner diagnostics.

\section{The Model}
\label{sec:ch5:model}

We work with the $\SU(2)$ BF spin foam amplitude. As with any spin foam model, it is built by contracting vertex amplitudes $\mathcal{A}_v$, edge amplitudes $\mathcal{A}_e$, and face amplitudes $\mathcal{A}_f$ along the 2-complex dual to a triangulation $\Delta$: vertices $v$, edges $e$, and faces $f$ correspond respectively to the 4-simplices, tetrahedra, and triangles of $\Delta$. Each face $f$ carries a spin $j_f$, and each edge $e$ an intertwiner $i_e$ among the four spins of the triangles bounding the corresponding tetrahedron. Only the non-boundary (bulk) spins and intertwiners are summed over. The boundary data are held fixed by the chosen boundary state.

The vertex amplitude associated with a 4-simplex $v$ is
\begin{equation}
    \mathcal{A}_v(j_f,i_e) = \{15j\}(j_f,i_e)\,,
    \label{eq:vertex_amplitude}
\end{equation}
the $\SU(2)$ $\{15j\}$-symbol \cite{Varshalovich:1988vb,Yutsis:1962wm}, obtained by contracting the five 4-valent intertwiners of the boundary tetrahedra of $v$ through their ten shared triangles (see App.~\ref{app:bf} for the explicit definition). This is the defining difference with the EPRL-FK vertex: rather than a superposition of $\{15j\}$-symbols weighted by $\gamma$-dependent booster functions enforcing the simplicity constraint \cite{Speziale:2016axj,Dona:2019dkf,Dona:2022dxs}, the $\SU(2)$ BF vertex amplitude \eqref{eq:vertex_amplitude} \emph{is} a single $\{15j\}$-symbol, with no auxiliary sum and no dependence on the Immirzi parameter.

We use the graphical recoupling basis, whose intertwiner of label $i$ has squared norm $1/d_i$, with $d_i=2i+1$ (see \cref{eq:intertwiner_basis_norm}). In this convention, $\{15j\}$ in \cref{eq:vertex_amplitude} denotes the contraction of five such intertwiners. Each internal edge $e$, shared by exactly two vertices, carries $\mathcal{A}_e(i_e)=d_{i_e}$, the inverse squared norm required by the resolution of the identity. The face weight $\mathcal{A}_f(j_f)=d_{j_f}$, where $d_{j_f}\equiv 2j_f+1$ is the representation dimension, is likewise fixed by requiring the amplitude to satisfy the correct composition rules under gluing of triangulations \cite{Bianchi:2010fj}. The transition amplitude on $\Delta$ is
\begin{equation}
    \mathcal{A}(\Delta) = \sum_{\{j_f\}} \sum_{\{i_e\}} \prod_f \mathcal{A}_f(j_f) \prod_e \mathcal{A}_e(i_e) \prod_v \mathcal{A}_v(j_f,i_e)\,,
    \label{eq:amplitude_definition}
\end{equation}
with the sums restricted to bulk spins and intertwiners as above. We use the amplitude \eqref{eq:amplitude_definition} throughout the numerical computation. We write it explicitly for the triangulations considered here in \cref{sec:ch5:config}, and we detail its evaluation in the recoupling basis in App.~\ref{app:bf}.

Boundary data are specified by $\SU(2)$ spin networks on the boundary graph, compatible with the kinematical Hilbert space of loop quantum gravity. We use coherent boundary states so spins determine triangle areas and intertwiners peak on classical polyhedra. These states define the discrete classical geometry that characterizes the boundary.

At large spin, the amplitude \eqref{eq:amplitude_definition} oscillates and admits a stationary-phase interpretation in which contributions are selected by the boundary data \cite{Barrett:2009asymptotics}. Asymptotically, the sum over internal spins selects configurations that are flat, reflecting the trivial dynamics of BF theory: in this limit, the bulk geometry is entirely determined by the boundary state. In practice, we work at finite spins, where this correspondence only holds approximately. Part of the present analysis identifies the range of boundary data for which an approximately compatible bulk geometry still exists.

\section{Configuration and Observables}
\label{sec:ch5:config}

\begin{figure*}[t]
    \centering
    \refstepcounter{figure}\label{fig:symmetric_wiring}
    \edef\thesubfigure{\thefigure\noexpand\alph{subfigure}}
    \captionsetup[subfigure]{labelformat=simple,labelsep=period}
    \makeatletter\renewcommand{\p@subfigure}{}\makeatother
    \begin{subfigure}[b]{0.48\textwidth}
        \centering
        \includegraphics[width=\textwidth]{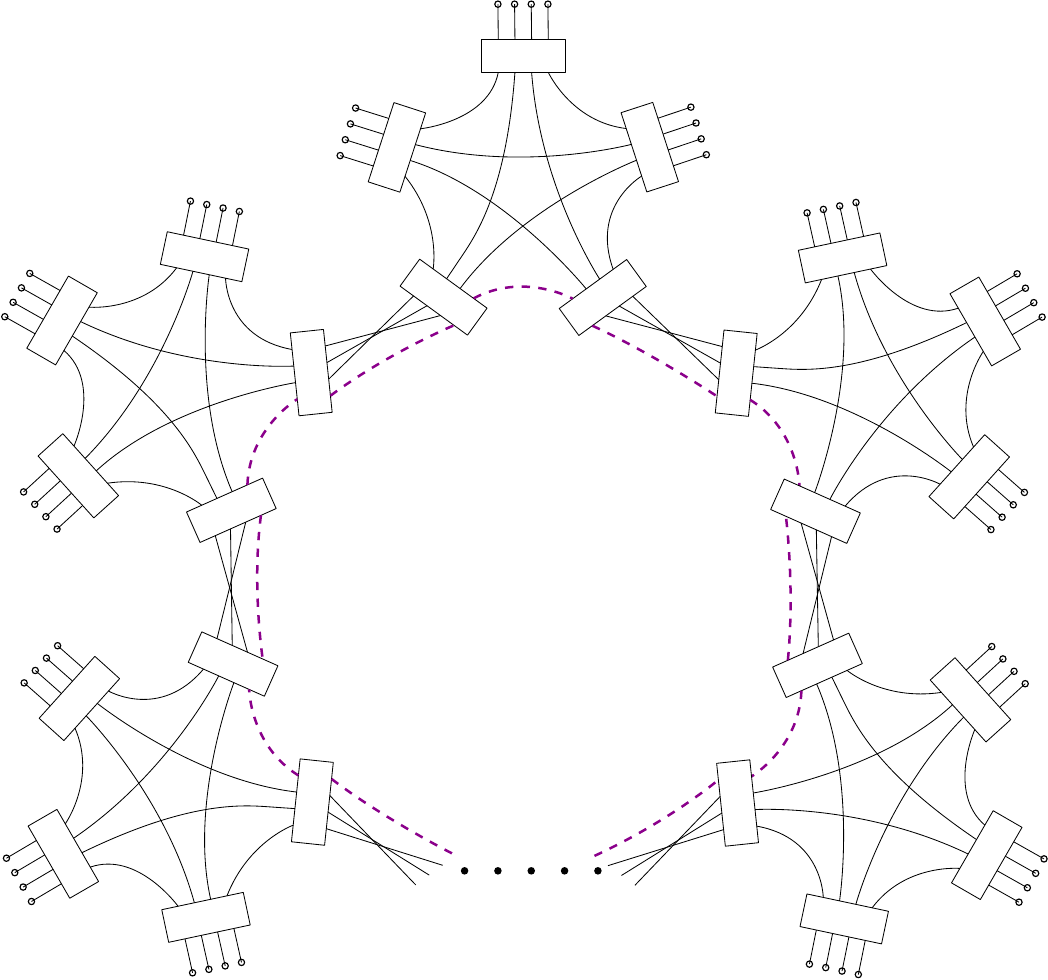}
        \caption{Wiring diagram of the $\Delta_n$ triangulation. Each line corresponds to a spinfoam face (triangle), and each box to an SU(2) integration (tetrahedron). The only bulk face is highlighted in red. Small circles represent the coherent states used in the boundary intertwiners.}
    \end{subfigure}
    \hfill
    \begin{subfigure}[b]{0.48\textwidth}
        \centering
        \includegraphics[width=.6\textwidth]{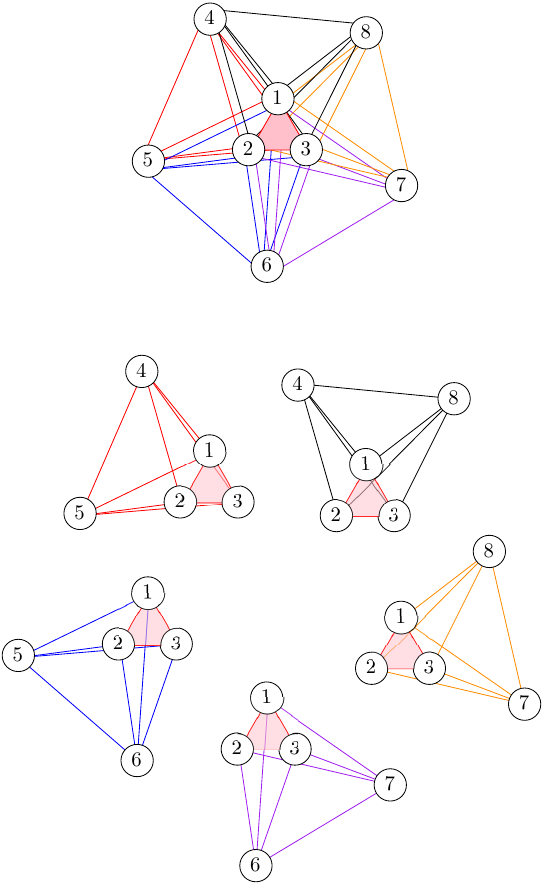}
        \caption{The $\Delta_5$ triangulation in terms of its geometric vertices. We shade the bulk triangle in red. The diagram makes the symmetry explicit, and the generalization to arbitrary $n$ is straightforward.}
    \end{subfigure}
\end{figure*}

We consider triangulations $\Delta_n$ constructed by gluing an odd number $n\geq3$ of 4-simplices around a common triangle. The numerical analysis uses $n\in\{3,5,7,9\}$. Each simplex has the same combinatorial structure, so that the complex is a symmetric composition of identical building blocks. The simplices are glued pairwise along tetrahedra containing the shared triangle, forming a closed network in which the number of simplices grows while the local structure remains fixed (\cref{fig:symmetric_wiring} shows an example with $n=5$). We take all boundary spins to be equal to $j$, fixed within each calculation, with $j \in \{20, 22, 24\}$ throughout. The only bulk spin is the one associated with the shared triangle, denoted by $j_{\text{b}}$. The bulk intertwiners are associated with the internal edges: since each of the $n$ simplices contributes one internal tetrahedron, shared with its neighbor, there are $n$ bulk intertwiners in total, denoted $i_{\text{b}_1}, \dots, i_{\text{b}_n}$.

The amplitude on $\Delta_n$ is
\begin{equation}
\begin{aligned}
    \mathcal{A}_{\Delta_n}
    &= \sum_{j_{\text{b}}, i_{\text{b}_1}, \dots, i_{\text{b}_n}} d_{j_{\text{b}}} \\
    &\quad\times\prod_{e=1}^{n}d_{i_{\text{b}_e}}
    \prod_{v=1}^{n}\mathcal{A}_v(j_{\text{b}}, i_{\text{b}_v}, i_{\text{b}_{v+1}}).
\end{aligned}
\label{eq:amplitude_deltan}
\end{equation}
Here $\mathcal{A}_v(j_{\text{b}},a,b)$ denotes the graphical-basis vertex contribution after contraction with the coherent states on its three boundary tetrahedra. The common boundary spin $j$ and the coherent-state normals are held fixed in each calculation and omitted from this notation. The bulk intertwiner indices are cyclic, with $i_{\text{b}_{n+1}} \equiv i_{\text{b}_1}$.

The boundary geometry is specified by spin-network states with coherent intertwiners, corresponding to a consistent assignment of areas and shapes for a given classical geometry \cite{Livine:2007vk}. We choose the boundary of $n$ 4-simplices glued cyclically around a common triangle $f$, imposing cyclic symmetry around this triangle. Within each 4-simplex $v$, two tetrahedra meet at $f$. The angle between them is the four-dimensional dihedral angle $\alpha_v$, measured inside the 4-simplex in the two-dimensional plane orthogonal to $f$. After gluing, the deficit angle at $f$ is
$$
    \delta = 2\pi - \sum_{v\supset f}\alpha_v\,.
$$
This angle measures the curvature concentrated at the shared triangle: for $\delta=0$, the angular sectors fit together to form a full $2\pi$ turn, giving a locally flat geometry around $f$. Cyclic symmetry makes all the dihedral angles equal, $\alpha_v=\alpha$, so prescribing the deficit angle fixes
$$
    \alpha = \frac{2\pi-\delta}{n}\,.
$$
Each 4-simplex is then completely determined by its nine boundary triangle areas, the imposed symmetry, and this internal dihedral angle.

From this classical geometry, we reconstruct the outward-pointing 3D unit normals to the triangular faces of each of the three boundary tetrahedra of every 4-simplex, in a common time-gauge convention. Together with the corresponding areas, these normals define Livine--Speziale (LS) coherent intertwiners \cite{Livine:2007vk}. Contracting these intertwiners with the corresponding legs of the vertex amplitude fixes the boundary data entering the numerical computation, following the same construction used in \cite{Dona:2019dkf}. App.~\ref{app:geometry} gives further details of the geometric reconstruction.

We define $a(j_{\text{b}})$ as the partial amplitude at fixed bulk spin, obtained from \cref{eq:amplitude_deltan} after summing over the internal intertwiners,
\begin{equation}
\begin{aligned}
    a(j_{\text{b}})
    &= d_{j_{\text{b}}}\sum_{i_{\text{b}_1},\dots,i_{\text{b}_n}} \\
    &\quad\times\prod_{v=1}^{n}
    \sqrt{d_{i_{\text{b}_v}}d_{i_{\text{b}_{v+1}}}}\,
    \mathcal{A}_v(j_{\text{b}}, i_{\text{b}_v}, i_{\text{b}_{v+1}}).
\end{aligned}
\label{eq:partial_amplitude}
\end{equation}
so that $\mathcal{A}_{\Delta_n} = \sum_{j_{\text{b}}} a(j_{\text{b}})$; this is the object used below to define the bulk spin expectation value. The square-root factors distribute each graphical edge weight equally between the two adjacent vertices.

The bulk-spin observable is the expectation value, in the path-integral sense,
\begin{equation}
    \expval{j_{\text{b}}} = \frac{\sum_{j_{\text{b}}} j_{\text{b}} \, a(j_{\text{b}})}{\sum_{j_{\text{b}}} a(j_{\text{b}})}\,.
\end{equation}
For the numerical plots, we use the relative dispersion (when the total amplitude and $\expval{j_{\text{b}}}$ are nonzero)
\begin{equation}
    D_j = \sqrt{\left|\frac{\expval{j_{\text{b}}^2}}{\expval{j_{\text{b}}}^2}-1\right|}\,,
\end{equation}
where
\begin{equation}
    \expval{j_{\text{b}}^2} = \frac{\sum_{j_{\text{b}}} j_{\text{b}}^2 \, a(j_{\text{b}})}{\sum_{j_{\text{b}}} a(j_{\text{b}})}\,.
\end{equation}
The quantity $D_j$ characterizes amplitude-weighted moments. The analogous intertwiner diagnostic $D_i$ is defined in \cref{app:numerics:sec:dispersion}. Since the weights can be complex, the moments and their central second moments can also be complex; when real, the latter can have either sign. Phase cancellations can reduce the dispersions even when the contributing labels are spread out, so these quantities differ from probability standard deviations. The expectation-value plots show the real part of the corresponding ratio.

Agreement with the classical reference and the size of $D_j$ provide useful comparisons between finite-spin amplitudes. Relating these moment diagnostics to geometric localization also calls for information about the contributing spin support, phases, stationary configurations, and the conditioning of the normalizing amplitude.

The classical reference is $j_{\text{classical}}=jA_{\text{shared}}$, where $A_{\text{shared}}$ is the reconstructed shared-triangle area in units of the common boundary area (\cref{app:geometry}). For each triangulation, we evaluate $\expval{j_{\text{b}}}/j_{\text{classical}}$ and $D_j$ over a deficit-angle grid spanning $[0,\pi]$. This compares the response near the flat configuration and at nonzero curvature for the different spin scales and numbers of simplices.


\section{Numerical Implementation}
\label{sec:ch5:numerics}
We exploit the symmetry of the triangulations to simplify the calculation. A direct sum over the $n$ internal intertwiner labels contains a number of terms that grows exponentially with $n$. Cyclic symmetry reduces this contraction to a matrix trace and requires only one coherent vertex matrix for each fixed set of parameters $(n,\delta,j,j_{\text{b}})$. We evaluate the vertex using \texttt{sl2cfoam-next} \cite{Gozzini:2021kbt}.
\subsection{Practical Implementation}
At fixed bulk spin $j_{\text{b}}$, the calculation produces a matrix with two free indices, one for each internal tetrahedron adjacent to a vertex. We obtain this matrix by contracting the other three vertex indices with the boundary coherent intertwiners. We first define the coefficient vectors used in that contraction.

For one boundary tetrahedron, let $\vec n=(\vec n^{(1)},\allowbreak\ldots,\allowbreak\vec n^{(4)})$ be its four outward unit face normals. All four face spins equal the boundary spin $j$. To each normal we associate a normalized spin coherent state $\ket{j,\vec n^{(\ell)}}$. The Livine--Speziale intertwiner $\ket{\iota(\vec n)}$ is the $\SU(2)$ group average of the tensor product of these four states. Expanding it in an orthonormal recoupling basis $\ket{i}_{\mathrm{N}}$ gives
\begin{equation}
\ket{\iota(\vec n)}=\sum_i c_i(\vec n)\ket{i}_{\mathrm{N}},
\label{eq:coherent_basis_expansion}
\end{equation}
where $i\in\{0,1,\ldots,2j\}$ runs over the admissible intermediate spins. Since every basis state is already invariant, its coefficient can be computed directly from the tensor product before group averaging:
\begin{equation}
c_i(\vec n)={}_{\mathrm{N}}\!\bra{i}
\left(\bigotimes_{\ell=1}^{4}\ket{j,\vec n^{(\ell)}}\right).
\label{eq:coherent_basis_coefficients}
\end{equation}
Thus each tetrahedron supplies a vector of complex numbers $c_i(\vec n)$, one per allowed recoupling label. This is the vector returned by \texttt{coherentstate\_\allowbreak{}compute}. Orthonormality refers to the basis states: the group-averaged coherent intertwiner itself need not have unit norm, and $\|\iota\|^2=\sum_i|c_i|^2$. An overall normalization of each fixed boundary state cancels from the normalized observable ratios.

The vertex routine uses the same orthonormal intertwiner basis as the coherent coefficients. Appendix~\ref{app:bf} records its relation to the graphical convention. One internal leg is evaluated in a different coupling channel from the one used for gluing. We denote its label in the routine's channel by $i'$ and the desired label by $i$. The change of basis is
\begin{equation}
R_{i'i}=(-1)^{j_{\text{b}}}\sqrt{d_{i'}d_i}
\left\lbrace\begin{array}{ccc}
j_{\text{b}} & j & i'\\ j & j & i
\end{array}\right\rbrace,
\label{eq:normalized_recoupling}
\end{equation}
with the phase fixed by the leg ordering and orientations in \cref{fig:symmetric_wiring}. We define $\widehat{\mathcal{A}}_v$ as the vertex tensor in the orthonormal gluing basis, obtained by applying $R$ to the routine's internal index in the different coupling channel. The full admissible-range transformation satisfies $\sum_{i'}R_{i'i}R_{i'k}=\delta_{ik}$, where $k$ is another label in the gluing channel.

For the three boundary tetrahedra, use the vectors $c_{i_k}(\vec n_k)$, $k=1,2,3$, obtained from the normals reconstructed in \cref{app:geometry}. Denote the two internal labels by $a\equiv i_{\text{b}_v}$ and $b\equiv i_{\text{b}_{v+1}}$. The matrix used in the numerical calculation is then
\begin{equation}
\begin{aligned}
M(j_{\text{b}})_{ab}
&=\sum_{i_1,i_2,i_3}
c_{i_1}(\vec n_1)c_{i_2}(\vec n_2)c_{i_3}(\vec n_3)
\\
&\quad\times\widehat{\mathcal{A}}_v(j_f,i_1,i_2,i_3,a,b).
\end{aligned}
\label{eq:coherent_vertex_normalized}
\end{equation}
Each coefficient vector contracts one boundary index of the vertex tensor, leaving $a$ and $b$ free. The contraction is multilinear in the boundary-ket coefficients. The dimension factors are already included in the orthonormal-basis vertex and coefficient vectors.

Because the boundary data and the gluing pattern are symmetric under cyclic permutation of the $n$ simplices, the same three boundary coherent states are contracted at every vertex, so $M(j_{\text{b}})$ is identical for all $v$. The orthonormal contraction already accounts for the graphical edge weights in \cref{eq:partial_amplitude}. The sum over the $n$ internal intertwiners, with the cyclic identification $i_{\text{b}_{n+1}} \equiv i_{\text{b}_1}$, then collapses to a single trace,
\[
a(j_{\text{b}}) = d_{j_{\text{b}}}\mathrm{Tr}\!\left[M(j_{\text{b}})^n\right].
\]
This is the key simplification: rather than evaluating $n$ distinct vertex amplitudes and summing over $n$ independent intertwiners, we only ever need the single matrix $M(j_{\text{b}})$, computed once per value of $j_{\text{b}}$ for fixed $(n,\delta,j)$. The matrix is the same at every vertex, but its coherent boundary data depend on $n$ and $\delta$ through $\alpha=(2\pi-\delta)/n$.

With the vertex and recoupling phases retained in $M(j_{\text{b}})$, the full amplitude is obtained by summing over all admissible bulk spins,
\[
\mathcal{A}_{\Delta_n} = \sum_{j_{\text{b}}} d_{j_{\text{b}}}\mathrm{Tr}\!\left[M(j_{\text{b}})^n\right].
\]

The admissible values of $j_{\text{b}}$ are determined by the $\SU(2)$ admissibility conditions of the adjacent four-valent intertwiners. Since three of the four spins entering each intertwiner are fixed to the boundary value $j$, the remaining spin is restricted to the finite range
\[
j_{\text{b}}\in\{0,1,\ldots,3j\}.
\]

We work with the intertwiner degrees of freedom in the $12$-recoupling basis described in App.~\ref{app:intertwiner}. In this basis, a four-valent intertwiner is represented by an intermediate spin label $i_{\text{b}}$. For fixed $j_{\text{b}}$, the admissible values of $i_{\text{b}}$ are determined by two sets of $\SU(2)$ triangle inequalities. The first arises from coupling two boundary spins,
\[
0\le i_{\text{b}}\le 2j,
\]
while the second arises from coupling the remaining boundary spin to the bulk spin,
\[
|j-j_{\text{b}}|
\le i_{\text{b}} \le
j+j_{\text{b}}.
\]
The allowed intertwiner values are therefore given by the intersection of these two ranges,
\[
i_{\text{b}}\in
\Bigl\{
|j-j_{\text{b}}|,
|j-j_{\text{b}}|+1,
\dots,
\min(2j,j+j_{\text{b}})
\Bigr\}.
\]

\subsection{Restricting the Amplitude}
\label{ch5:sec:numerics:subsec:restricting}

To probe the influence of the intertwiner sector, we restrict one recoupling channel to a finite window around its classical value and examine the resulting changes in bulk-spin observables and spin-sum cancellations. The unrestricted sum admits a wider class of configurations, potentially including non-Regge contributions such as vector geometries \cite{Barrett:2009asymptotics,Dona:2017dvf}.

The classical intertwiner value, denoted $i_{\text{classical}}$, is obtained from the reconstructed geometry through \cref{eq:intertwiner_external} (see also App.~\ref{app:intertwiner}) and is generally not an integer. The projected amplitude is obtained by restricting the intertwiner sum to
\[
i_{\text{b}} \in
\left[
\lfloor i_{\text{classical}}\rfloor -1,\,
\lfloor i_{\text{classical}}\rfloor +2
\right],
\]
where $\lfloor i_{\text{classical}} \rfloor$ denotes the greatest integer less than or equal to $i_{\text{classical}}$. The window is intersected with the admissible intertwiner range at each $j_{\text{b}}$. The recoupling label fixes one pairwise face-angle combination (see App.~\ref{app:intertwiner}). The window therefore controls one of the two classical tetrahedral shape degrees of freedom at fixed face areas \cite{Freidel:2010aa}, while its conjugate variable remains free.

The window contains at most four integer labels, and its span remains fixed as the spin scale increases. For fixed nondegenerate geometry, $i_{\text{classical}}$ scales as $j$ in the large-spin regime, so the relative window width shrinks as $1/j$. With signed or complex amplitude weights, the expectation value can lie outside the selected label interval.

The projector replaces the full intertwiner identity with a restricted sum, defining a modified BF amplitude. Comparing its bulk-spin moments with the unrestricted result tests how the selected shape channel affects the contributing terms and their interference. The topological and refinement properties of this modified amplitude require a separate study.

Several intertwiner restrictions were investigated as part of this study (\cref{app:numerics:sec:exploratory_projector_comparisons}). We use the finite-window projector as the reference prescription for the main analysis and report comparisons with alternative weights in the appendix.

We evaluate the projected amplitude using the same numerical procedure as the unrestricted one and compare the two throughout the analysis (\cref{sec:ch5:results}). The matrix form of the prescribed restriction is given in \cref{eq:projected_vertex_matrix,eq:projected_partial_amplitude}; projected observables use the same definitions with $a(j_{\text{b}})$ replaced by $a_P(j_{\text{b}})$.

\section{Results}
\label{sec:ch5:results}

The analysis proceeds in stages. First, we examine the bulk spin observables obtained from the unrestricted amplitude. We then analyze the intertwiner sector independently to assess its behavior under variations of the deficit angle, spin scale, and triangulation size. The unrestricted intertwiner moments can depart from their classical reference values despite the fixed boundary data; these departures concern amplitude-weighted diagnostics rather than a positive probability distribution. Motivated by this observation, we introduce an intertwiner projector and compare the resulting projected amplitudes with the unrestricted results. Throughout, we show the dependence on the deficit angle $\delta$ at fixed triangulation size $n$, with separate curves for different spin scales $j$.

\subsection{Bulk Spin Results (Unrestricted Amplitude)}
\label{sec:ch5:results:bulk_raw}

\begin{figure*}[tp]
\centering

\begin{subfigure}[b]{0.48\textwidth}
    \centering
    \includegraphics[width=\textwidth]
    {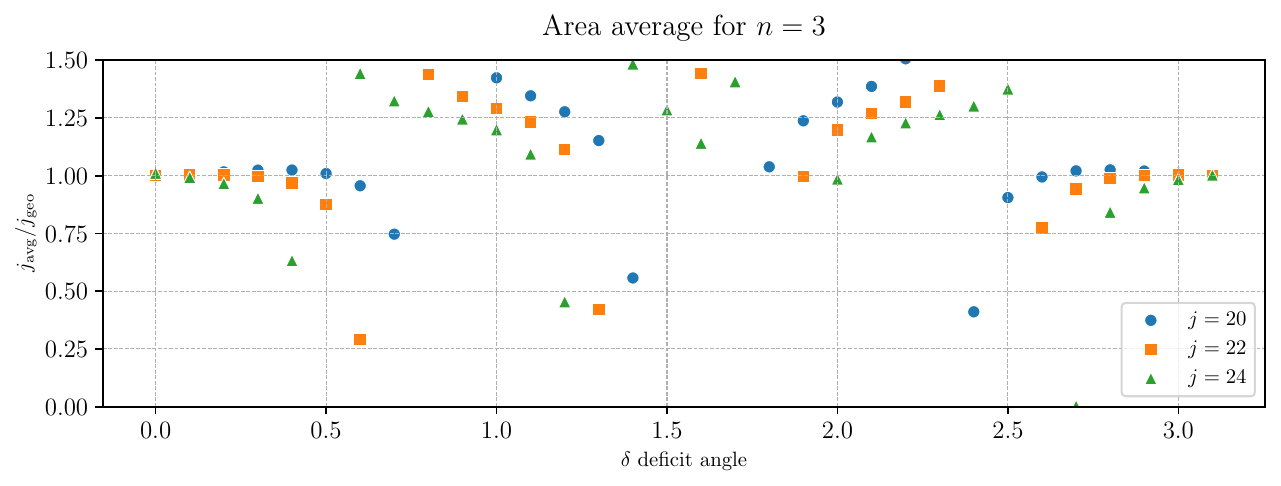}
    \caption{$n=3$}
\end{subfigure}
\hfill
\begin{subfigure}[b]{0.48\textwidth}
    \centering
    \includegraphics[width=\textwidth]
    {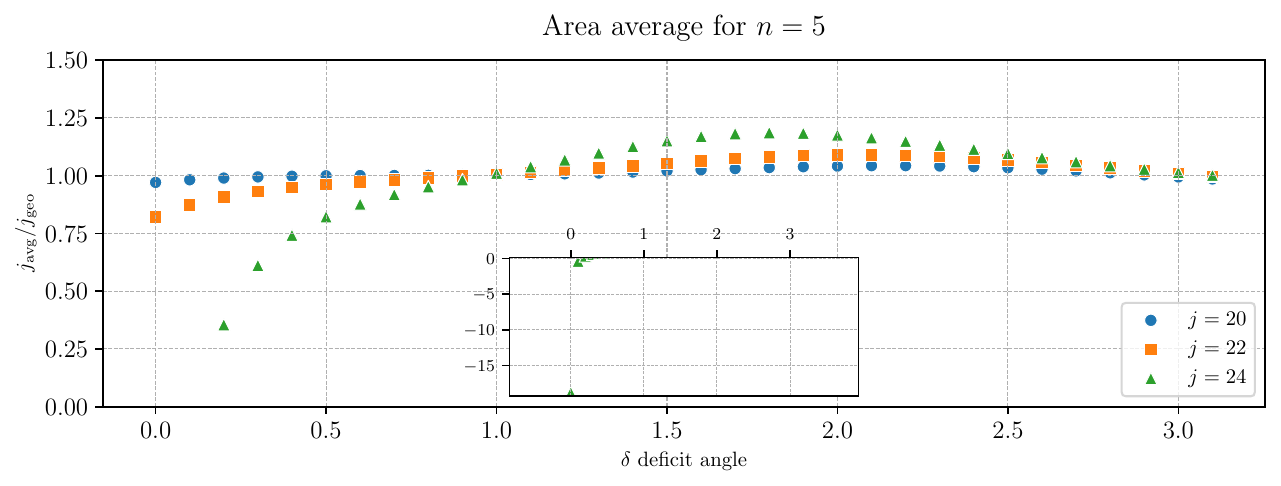}
    \caption{$n=5$, with negative values shown in the inset}
\end{subfigure}

\par\medskip

\begin{subfigure}[b]{0.48\textwidth}
    \centering
    \includegraphics[width=\textwidth]
    {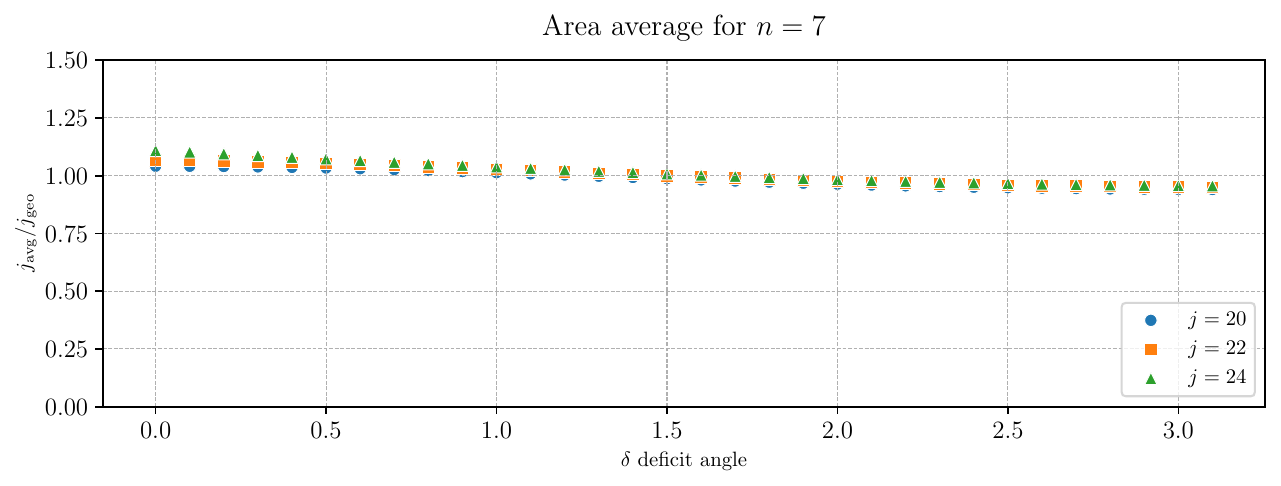}
    \caption{$n=7$}
\end{subfigure}
\hfill
\begin{subfigure}[b]{0.48\textwidth}
    \centering
    \includegraphics[width=\textwidth]
    {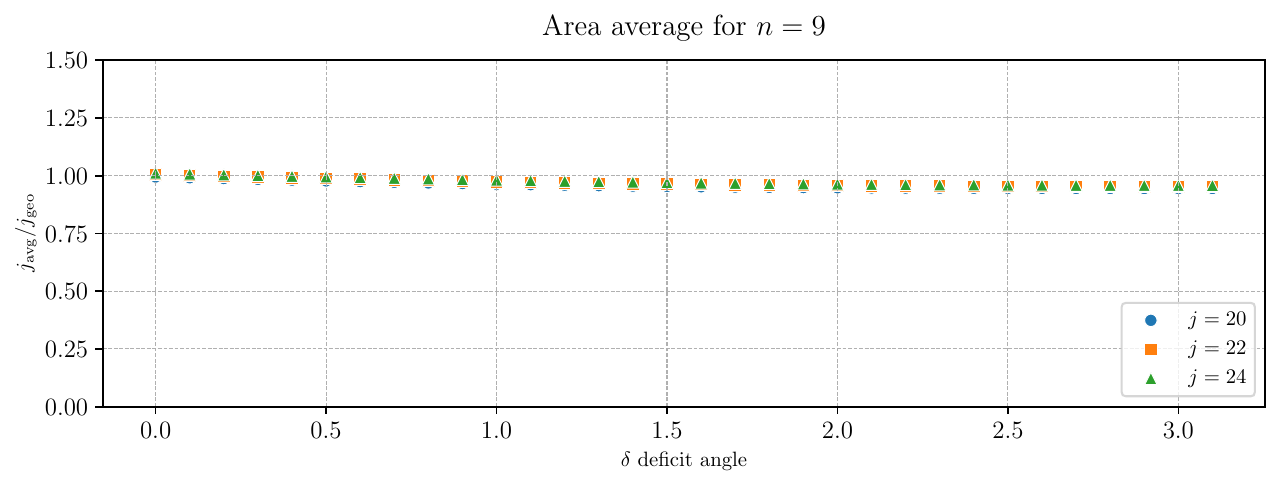}
    \caption{$n=9$}
\end{subfigure}

\caption{
Ratio $\operatorname{Re}\langle j_{\text{b}}\rangle/j_{\mathrm{classical}}$
as a function of the deficit angle $\delta$ for the triangulations
$\Delta_3$, $\Delta_5$, $\Delta_7$, and $\Delta_9$.
Each panel shows the results for $j=20,22,24$.
}
\label{fig:bulk_ratio_all_full}
\end{figure*}

\begin{figure*}[tp]
\centering

\begin{subfigure}[b]{0.48\textwidth}
    \centering
    \includegraphics[width=\textwidth]
    {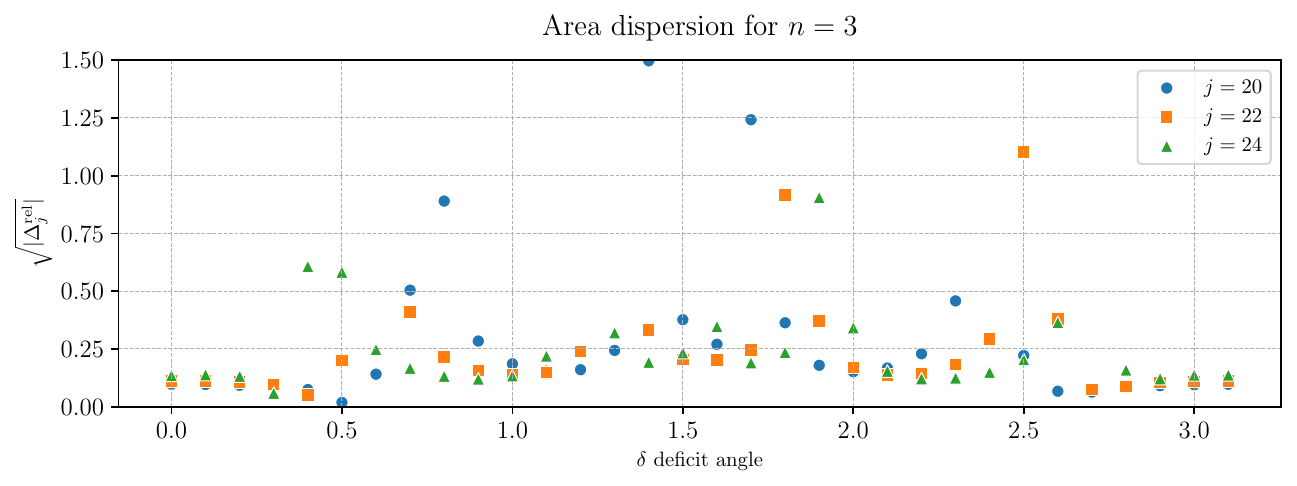}
    \caption{$n=3$}
\end{subfigure}
\hfill
\begin{subfigure}[b]{0.48\textwidth}
    \centering
    \includegraphics[width=\textwidth]
    {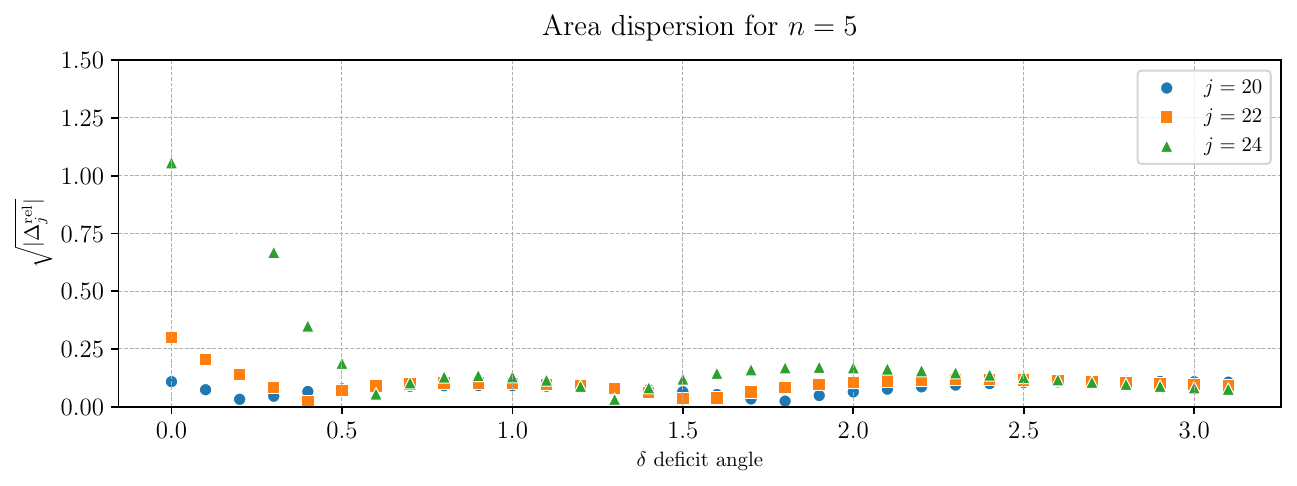}
    \caption{$n=5$}
\end{subfigure}

\par\medskip

\begin{subfigure}[b]{0.48\textwidth}
    \centering
    \includegraphics[width=\textwidth]
    {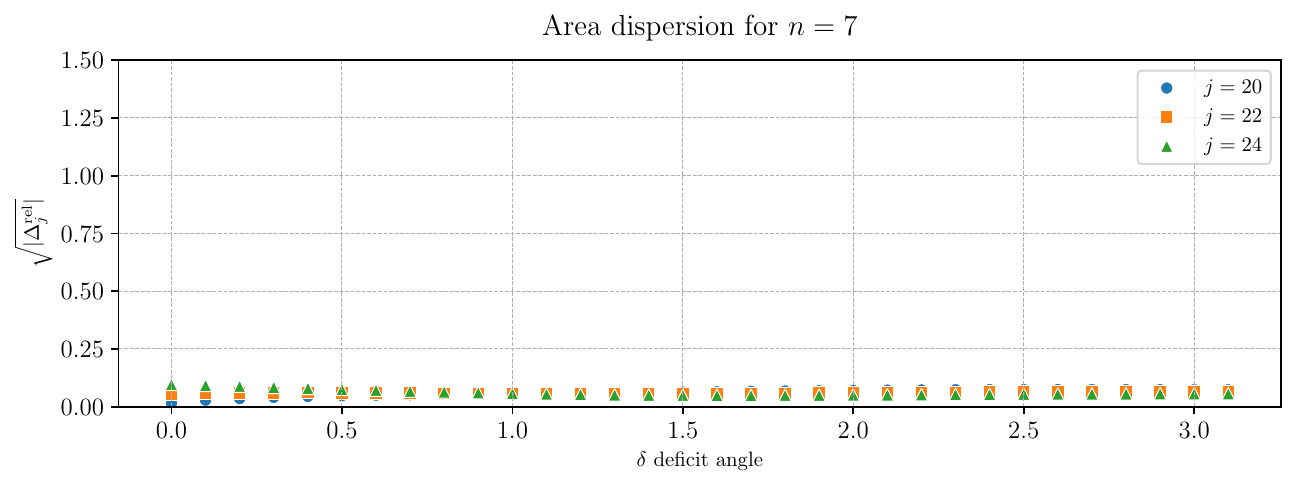}
    \caption{$n=7$}
\end{subfigure}
\hfill
\begin{subfigure}[b]{0.48\textwidth}
    \centering
    \includegraphics[width=\textwidth]
    {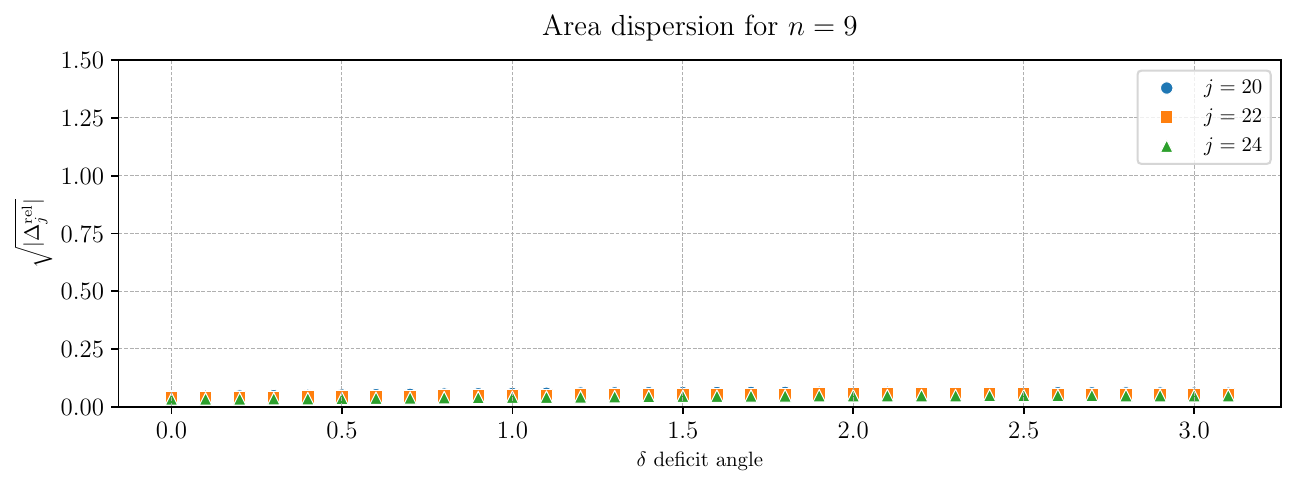}
    \caption{$n=9$}
\end{subfigure}

\caption{
Relative dispersion
$D_j=\sqrt{\left|\langle j_{\text{b}}^2\rangle/
\langle j_{\text{b}}\rangle^2 - 1\right|}$
of the bulk spin amplitude as a function of the deficit angle $\delta$
for the triangulations $\Delta_3$, $\Delta_5$, $\Delta_7$, and $\Delta_9$.
}
\label{fig:bulk_variance_all_full}
\end{figure*}

We begin by examining the bulk spin observables obtained from the unrestricted amplitude, which serve as the baseline for all subsequent comparisons. As discussed in \cref{sec:ch5:config}, we compare the bulk-spin expectation-value ratio with unity and track $D_j$ as amplitude-weighted diagnostics. These results provide a reference for evaluating the effects of intertwiner dynamics and of the projector introduced below. \cref{fig:bulk_ratio_all_full,fig:bulk_variance_all_full} show the ratio and its relative dispersion as functions of the deficit angle $\delta$ for $n=3,5,7,9$ and $j=20,22,24$.

Several features are apparent from the unrestricted results. Let us examine them in turn. For $n=3$, the expectation value remains in reasonable agreement with the classical prediction over a finite range of deficit angles $\delta \in [0,0.5]$. A clear trend emerges: as the spin scale increases from $j=20$ to $j=24$, the bulk spin deviates from its geometrical value at progressively smaller $\delta$. This trend describes a narrowing of the finite-spin range in which the expectation value follows the curved geometric target. Consequently, agreement with the classical geometry is expected to persist over a progressively narrower range of nonzero deficit angles. 

For $n=5$, a pronounced discrepancy appears at $\delta=0$, where the ratio $\expval{j_{\text{b}}}/j_{\text{classical}}$ becomes strongly negative. Such values are allowed for path-integral expectation values because the amplitude carries complex phases. A sharp spike in the relative dispersion accompanies this change. Away from the flat configuration, agreement improves substantially: the ratio generally remains close to unity and the relative dispersion returns to its typical small values.

For $n=7$, significant deviations at $\delta=0$ remain visible and are substantially larger than the estimated numerical uncertainties, although the negative values observed for $n=5$ are no longer present. The corresponding relative dispersion remains small overall but increases noticeably near the flat configuration.

For $n=9$, the ratio remains close to unity throughout the parameter range considered. The associated relative dispersion is smaller than that observed for $n=7$. 

Overall, the largest discrepancies occur near the flat configuration, whereas for nonzero deficit angles the unrestricted amplitudes typically yield ratios close to unity for $n>3$, with correspondingly small relative dispersions away from $\delta=0$. The spin-contribution and cumulative-sum diagnostics in \cref{sec:ch5:results:saddle} show large cancellations among spin contributions for the most anomalous case, $n=5$ at $\delta=0$. The cumulative sums make large excursions before reaching their final value. Such behavior can arise from oscillations within one stationary region or from interference between several critical configurations.

\subsection{Intertwiner Results (Unrestricted Amplitude)}
\label{sec:ch5:results:int_raw}

We next turn to the intertwiner degrees of freedom. With the boundary geometry and spins fixed, the intertwiner moments test the response of the remaining recoupling channel.

\begin{figure*}[tp]
\centering
\begin{subfigure}[b]{0.48\textwidth}
    \centering
    \includegraphics[width=\textwidth]
    {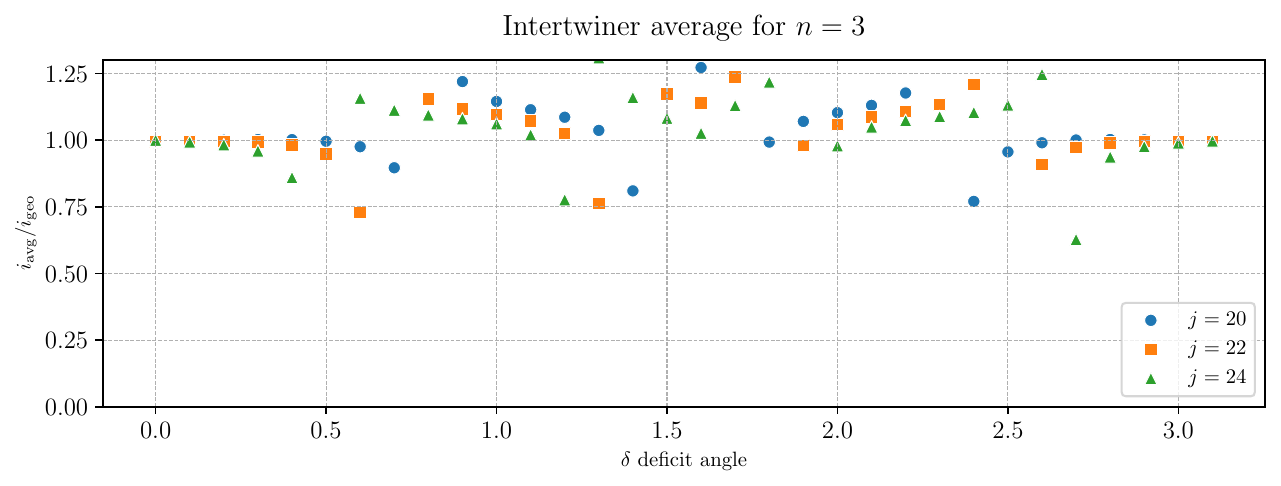}
    \caption{$n=3$}
\end{subfigure}
\hfill
\begin{subfigure}[b]{0.48\textwidth}
    \centering
    \includegraphics[width=\textwidth]
    {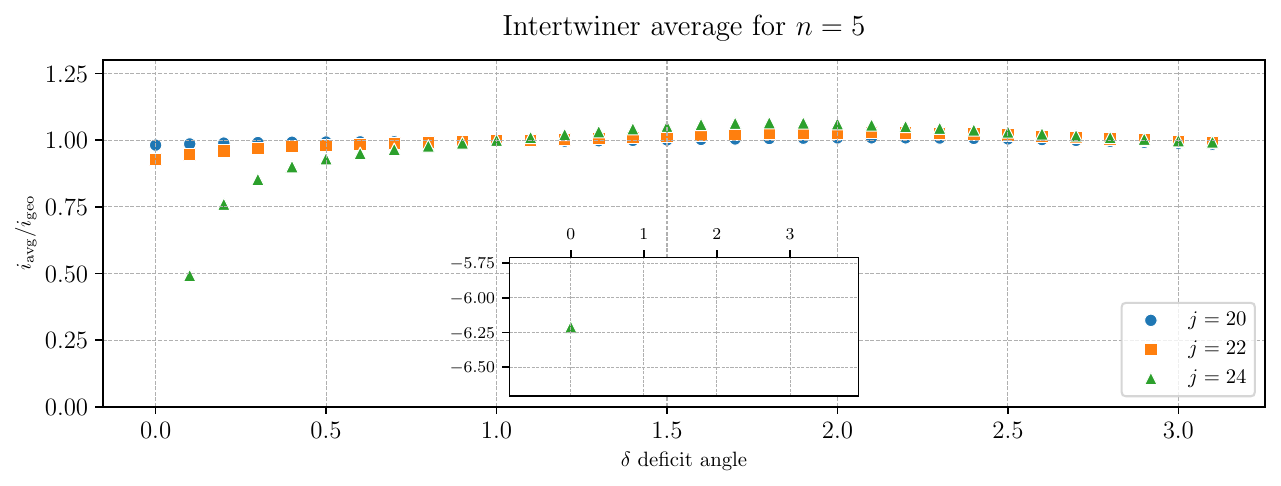}
    \caption{$n=5$}
\end{subfigure}

\par\medskip

\begin{subfigure}[b]{0.48\textwidth}
    \centering
    \includegraphics[width=\textwidth]
    {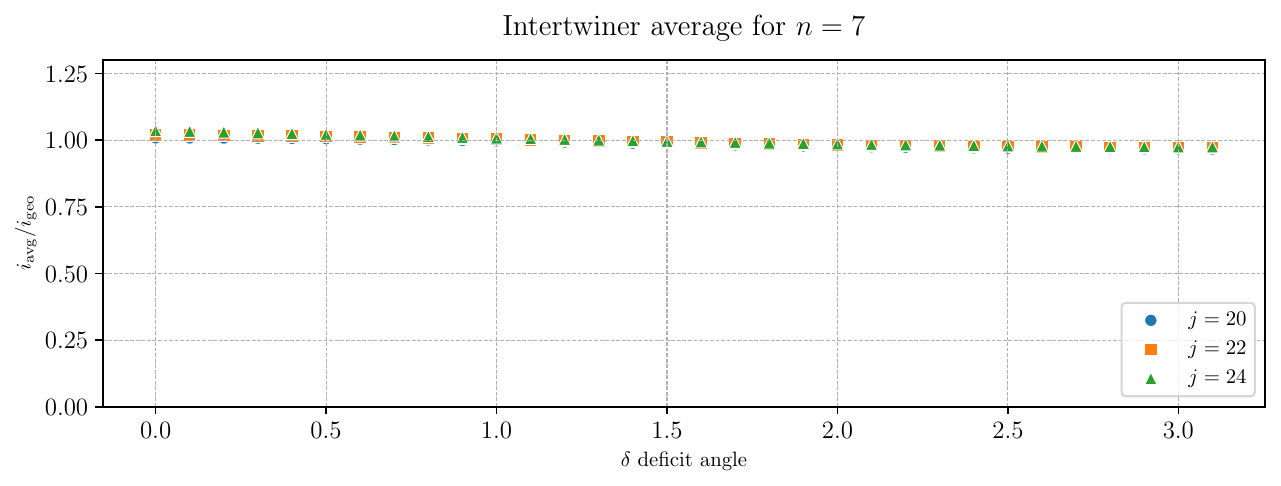}
    \caption{$n=7$}
\end{subfigure}
\hfill
\begin{subfigure}[b]{0.48\textwidth}
    \centering
    \includegraphics[width=\textwidth]
    {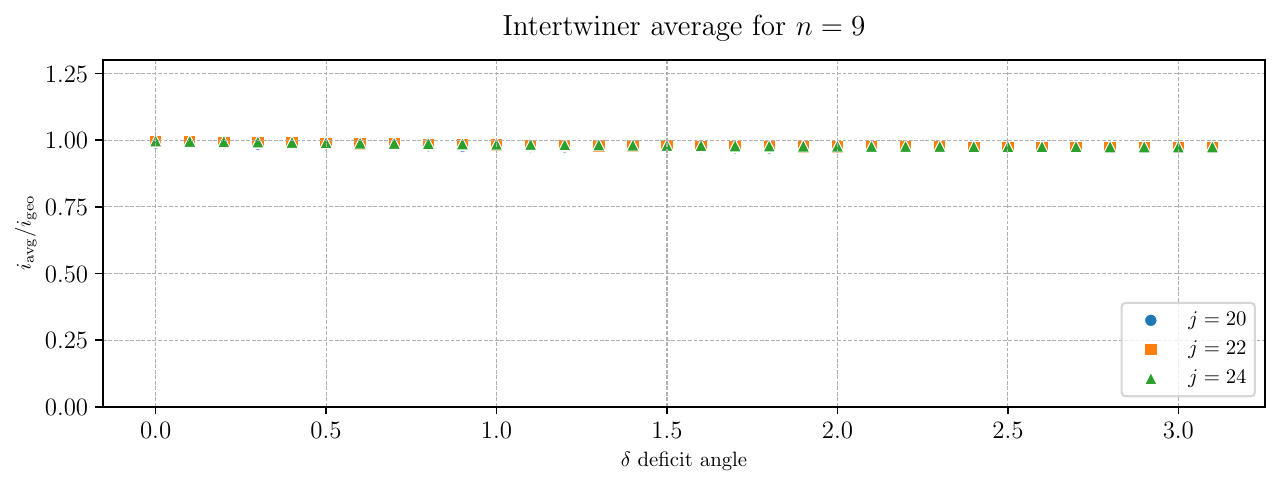}
    \caption{$n=9$}
\end{subfigure}

\caption{
Ratio $\operatorname{Re}\langle i_{\text{b}}\rangle/i_{\mathrm{classical}}$
as a function of the deficit angle $\delta$ for the triangulations
$\Delta_3$, $\Delta_5$, $\Delta_7$, and $\Delta_9$.
Each panel shows the results for $j=20,22,24$.
}
\label{fig:intertwiner_average_all_full}
\end{figure*}

\begin{figure*}[tp]
\centering

\begin{subfigure}[b]{0.48\textwidth}
    \centering
    \includegraphics[width=\textwidth]
    {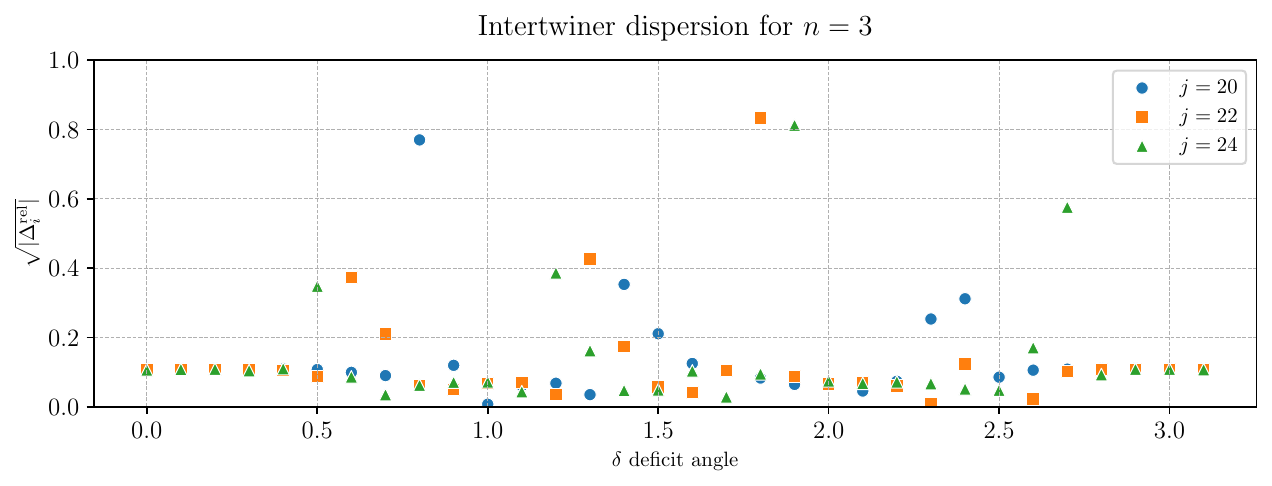}
    \caption{$n=3$}
\end{subfigure}
\hfill
\begin{subfigure}[b]{0.48\textwidth}
    \centering
    \includegraphics[width=\textwidth]
    {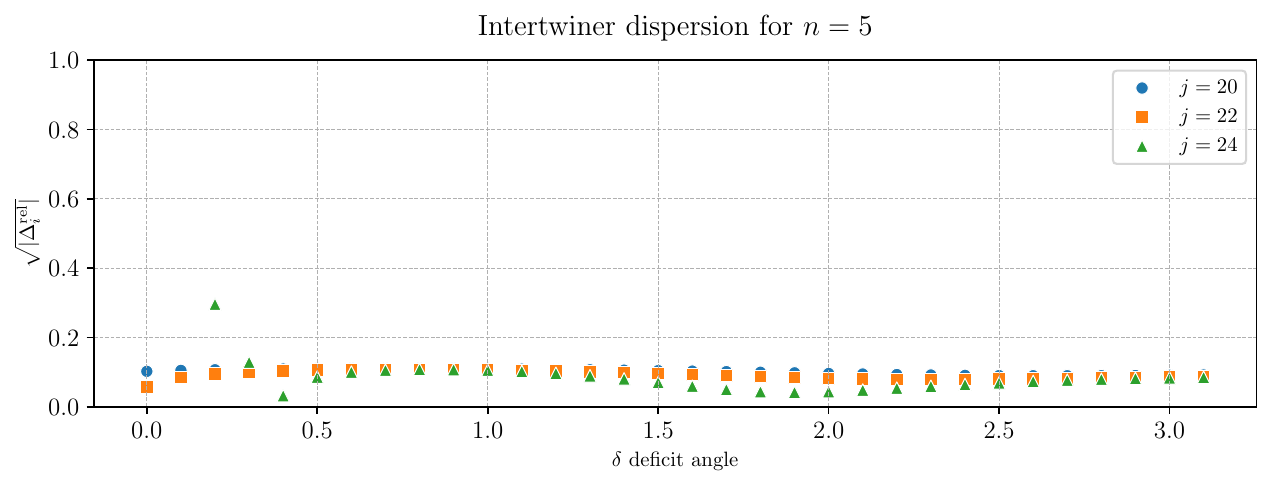}
    \caption{$n=5$}
\end{subfigure}

\par\medskip

\begin{subfigure}[b]{0.48\textwidth}
    \centering
    \includegraphics[width=\textwidth]
    {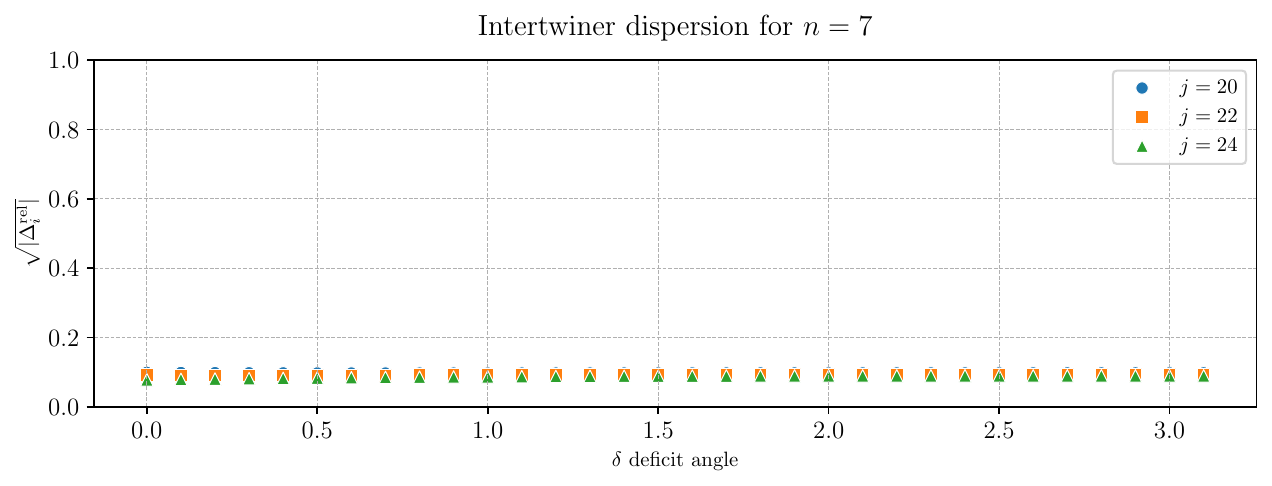}
    \caption{$n=7$}
\end{subfigure}
\hfill
\begin{subfigure}[b]{0.48\textwidth}
    \centering
    \includegraphics[width=\textwidth]
    {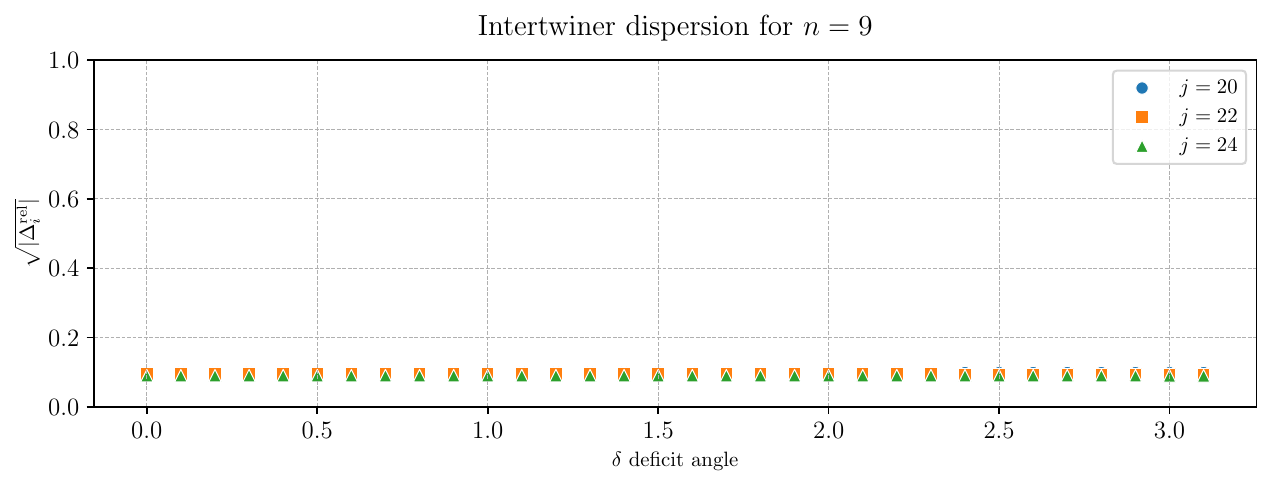}
    \caption{$n=9$}
\end{subfigure}

\caption{
Relative dispersion
$D_i=\sqrt{\left|\langle i_{\text{b}}^2\rangle/\langle i_{\text{b}}\rangle^2 - 1\right|}$
of the intertwiner amplitude as a function of the deficit angle $\delta$
for the triangulations $\Delta_3$, $\Delta_5$, $\Delta_7$, and $\Delta_9$.
}
\label{fig:intertwiner_variance_all_full}
\end{figure*}

We evaluate the moments of the intertwiner label on one internal tetrahedron using \cref{eq:intertwiner_moments}; cyclic symmetry makes the choice of tetrahedron immaterial. We compare its expectation value with the classical reference derived from the angle relation in App.~\ref{app:intertwiner}. \cref{fig:intertwiner_average_all_full,fig:intertwiner_variance_all_full} show the ratio $\langle i_{\text{b}}\rangle/i_{\text{classical}}$ and its relative dispersion as functions of the deficit angle $\delta$.

The intertwiner observables closely track the behavior found in the bulk spin sector. Their dependence on the deficit angle, spin scale, and triangulation size follows the same qualitative pattern: regions where the bulk spin ratio is close to its classical value are accompanied by an intertwiner ratio near unity and small relative dispersion, while anomalous regions appear in both sectors. In particular, the $n=5$, $\delta\approx0$ case, where the bulk spin ratio becomes strongly negative, is also anomalous in the intertwiner sector: the ratio $\langle i_{\text{b}}\rangle/i_{\text{classical}}$ drops well below unity and the relative dispersion increases sharply.

This correlation highlights the intertwiner as an active degree of freedom whose amplitude-weighted moments closely track the bulk-spin response. The cumulative-sum diagnostics below reveal large spin-sum cancellations in the $n=5$ flat case, motivating a comparison with the sum restricted around the classical intertwiner value.

\subsection{Bulk Spin Results (Projected Amplitude)}
\label{sec:ch5:results:bulk_proj}
\begin{figure*}[tp]
\centering

\begin{subfigure}[b]{0.48\textwidth}
    \centering
    \includegraphics[width=\textwidth]
    {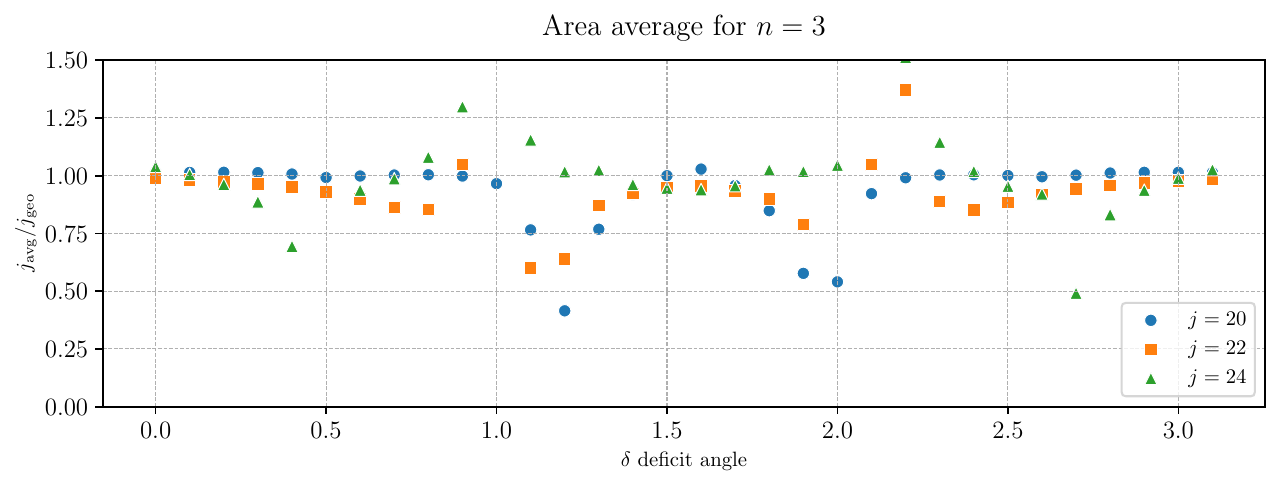}
    \caption{$n=3$}
\end{subfigure}
\hfill
\begin{subfigure}[b]{0.48\textwidth}
    \centering
    \includegraphics[width=\textwidth]
    {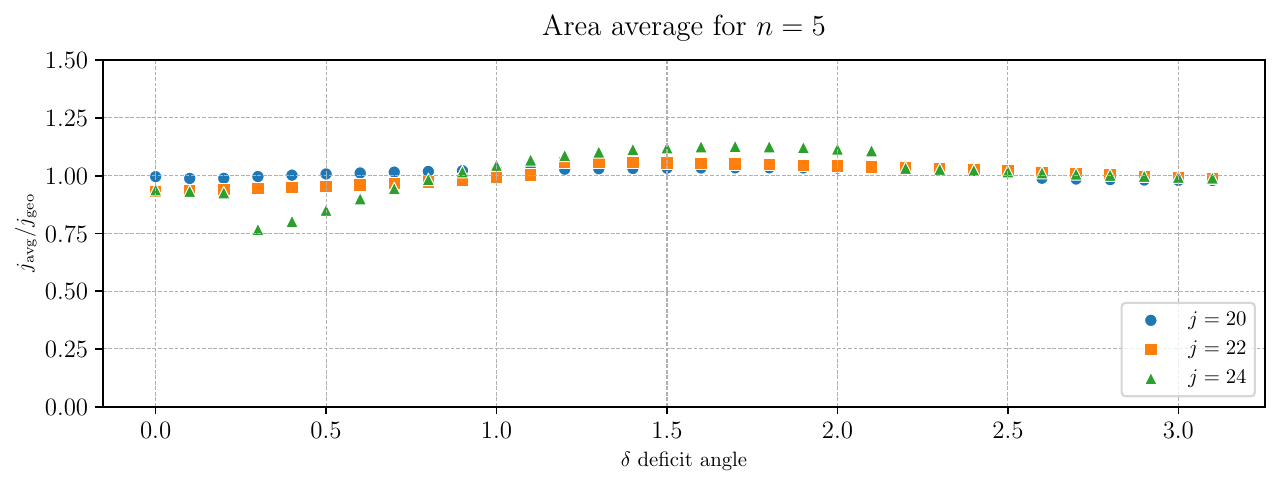}
    \caption{$n=5$}
\end{subfigure}

\par\medskip

\begin{subfigure}[b]{0.48\textwidth}
    \centering
    \includegraphics[width=\textwidth]
    {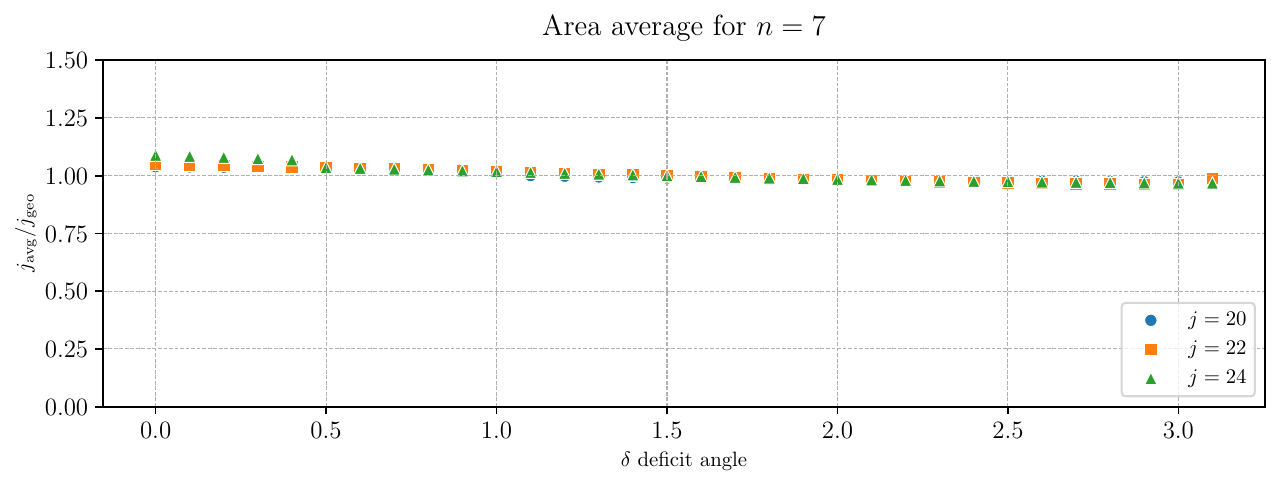}
    \caption{$n=7$}
\end{subfigure}
\hfill
\begin{subfigure}[b]{0.48\textwidth}
    \centering
    \includegraphics[width=\textwidth]
    {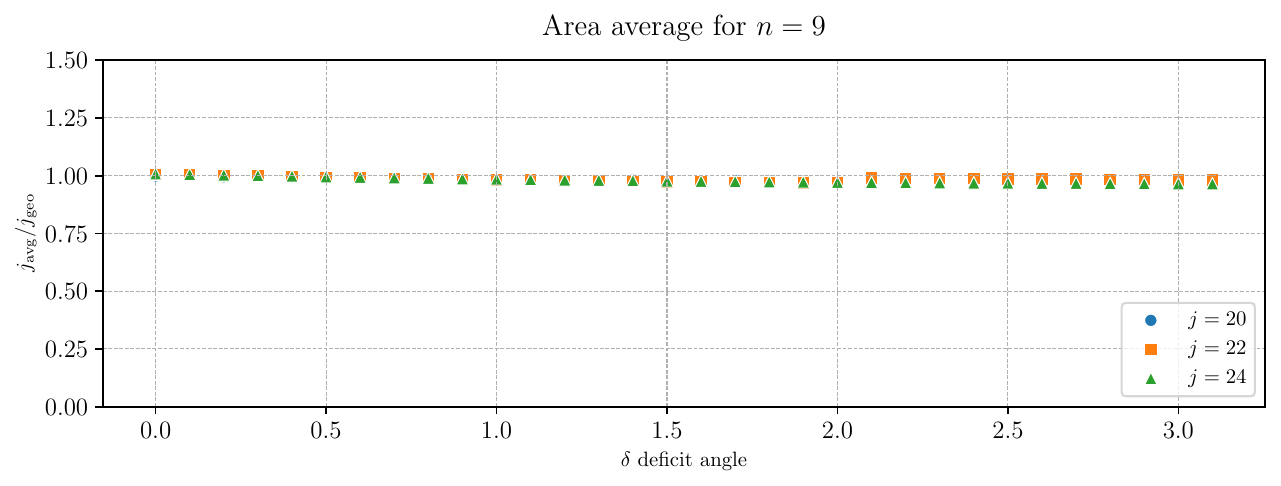}
    \caption{$n=9$}
\end{subfigure}

\caption{
Ratio $\operatorname{Re}\langle j_{\text{b}}\rangle/j_{\mathrm{classical}}$
as a function of the deficit angle $\delta$ for the triangulations
$\Delta_3$, $\Delta_5$, $\Delta_7$, and $\Delta_9$
after application of the intertwiner projector.
Each panel shows the results for $j=20,22,24$.
}
\label{fig:bulk_ratio_all_hard}
\end{figure*}

\begin{figure*}[tp]
\centering

\begin{subfigure}[b]{0.48\textwidth}
    \centering
    \includegraphics[width=\textwidth]
    {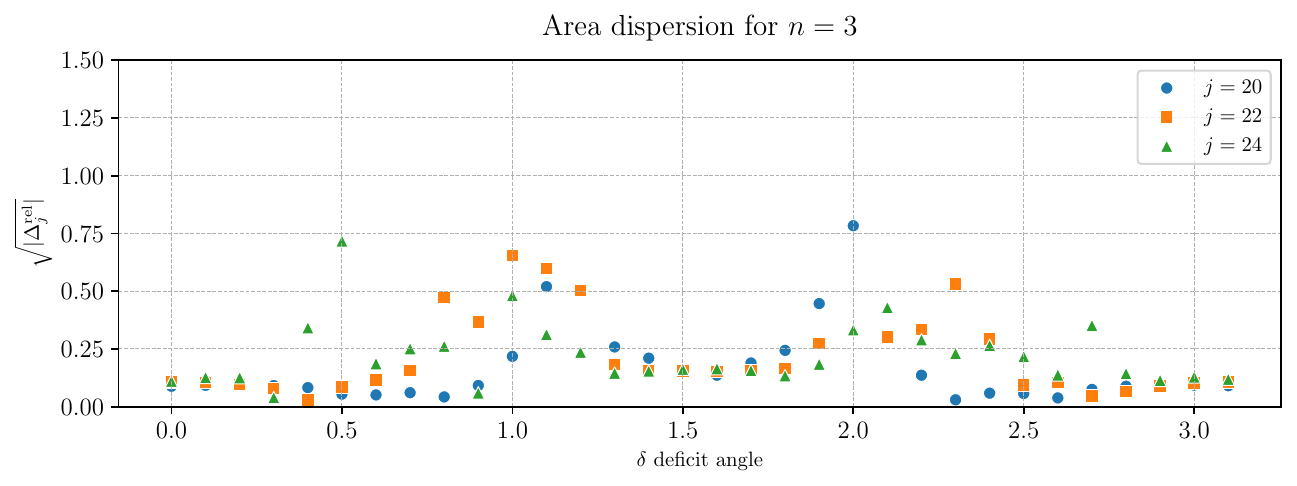}
    \caption{$n=3$}
\end{subfigure}
\hfill
\begin{subfigure}[b]{0.48\textwidth}
    \centering
    \includegraphics[width=\textwidth]
    {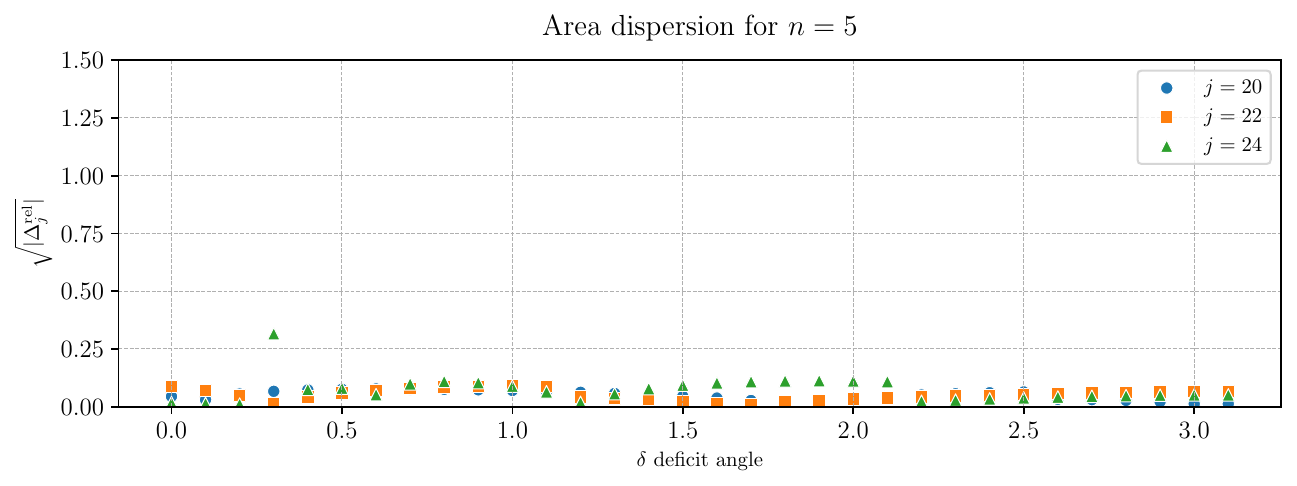}
    \caption{$n=5$}
\end{subfigure}

\par\medskip

\begin{subfigure}[b]{0.48\textwidth}
    \centering
    \includegraphics[width=\textwidth]
    {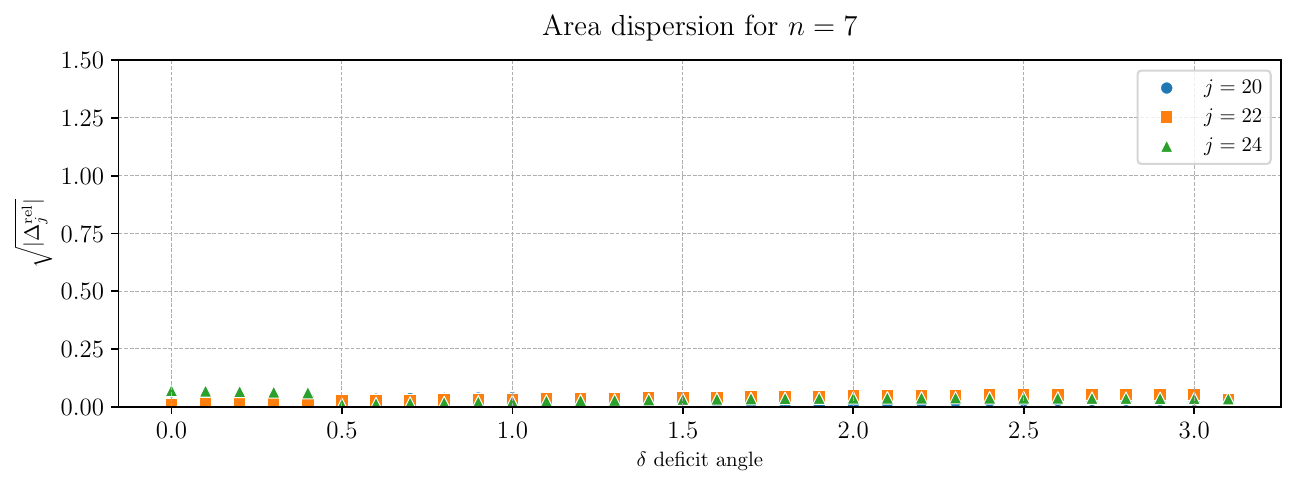}
    \caption{$n=7$}
\end{subfigure}
\hfill
\begin{subfigure}[b]{0.48\textwidth}
    \centering
    \includegraphics[width=\textwidth]
    {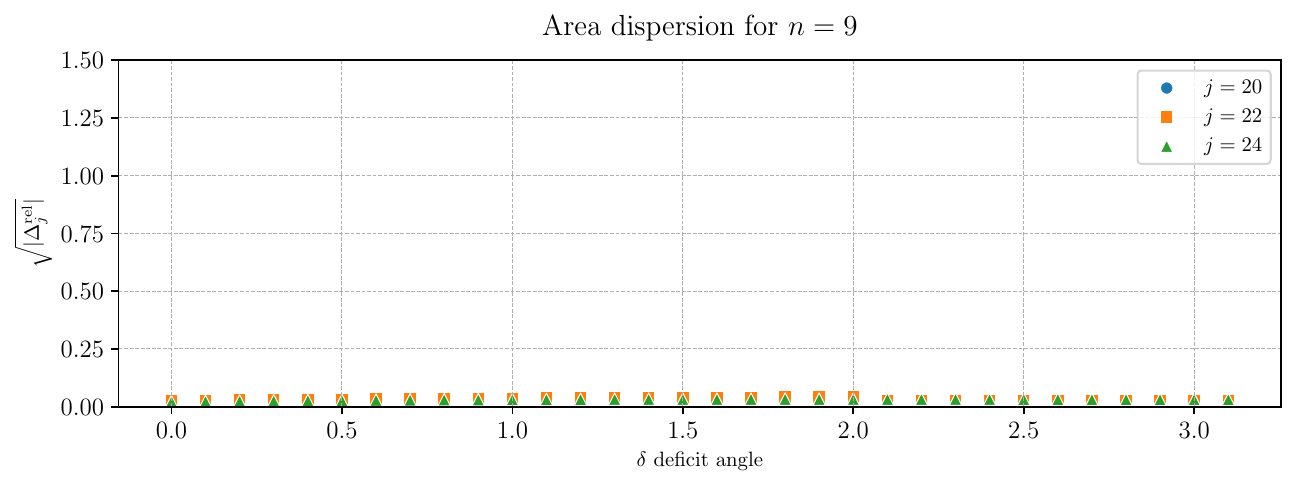}
    \caption{$n=9$}
\end{subfigure}

\caption{
Relative dispersion
$D_j=\sqrt{\left|\langle j_{\text{b}}^2\rangle/
\langle j_{\text{b}}\rangle^2 - 1\right|}$
of the bulk spin amplitude as a function of the deficit angle $\delta$
for the triangulations $\Delta_3$, $\Delta_5$, $\Delta_7$, and $\Delta_9$
after application of the intertwiner projector.
}
\label{fig:bulk_variance_all_hard}
\end{figure*}

We repeat the bulk spin analysis for the projected amplitude, obtained by restricting the intertwiner sum as described in \cref{sec:ch5:numerics}. Comparing these results with the unrestricted case tests how strongly the bulk-spin observables and cancellation diagnostics respond to the internal-intertwiner restriction. \cref{fig:bulk_ratio_all_hard,fig:bulk_variance_all_hard} show the corresponding ratio and relative dispersion after applying the projector.

The projected amplitude modifies the bulk spin observables in a way that depends on the triangulation. For $n=3$, the qualitative behavior remains similar to the unrestricted case, but deviations from the classical value set in at larger deficit angles for all spin values considered. For $n=5$, the negative ratios observed at $\delta=0$ disappear entirely, and the discrepancy with the geometric prediction drops to a few percent. For $n=7$, by contrast, the projected results are essentially indistinguishable from the unrestricted ones, indicating that restricting the intertwiner sector does not remove the mild deviation already present at $\delta=0$. For $n=9$, the ratio remains close to unity throughout the parameter range considered, as in the unrestricted analysis.

The projected intertwiner ratios remain close to their classical references in most cases, with small relative dispersions and the largest residual deviations for $n=3$ (see \cref{app:numerics:sec:projected_intertwiner_results}). This behavior is consistent with the prescribed restriction near the classical intertwiner value. The bulk-spin response provides a complementary result: for $n=5$, the modified amplitude removes the negative value at the nominally flat configuration and substantially improves agreement with the geometric reference.

Some projected observables change abruptly between $\delta=0$ and $\delta=0.5$. This is a discretization effect of the finite-window projector, not a discontinuity in the underlying amplitude. As the classical intertwiner value varies with the geometry, the center of the projection window can cross an integer boundary and shift by one intertwiner unit. Since the projector acts on a discrete set of intertwiner labels, this shift changes the configurations included in the projected sum and can produce a corresponding jump in the expectation values. Explicit checks confirm that the effect is due to the motion of the projection window rather than a qualitative change in the amplitude itself.

\subsection{Spin-Contribution and Cumulative-Sum Analysis} 
\label{sec:ch5:results:saddle}

To resolve how individual spin contributions combine into the final amplitude, we examine normalized contributions and cumulative sums, following the numerical strategy of \cite{Dona:2019jab,Dona:2020tvv}. For fixed boundary data and a nonzero total amplitude, define
\begin{equation}
\rho(j_{\text{b}})=\frac{a(j_{\text{b}})}{\mathcal{A}_{\Delta_n}},\qquad
\mathcal{A}_{\Delta_n}=\sum_{j_{\text{b}}=0}^{3j}a(j_{\text{b}}).
\label{eq:normalized_spin_contribution}
\end{equation}
The normalized cumulative sum up to an integer cutoff $J\in\{0,\ldots,3j\}$ is
\begin{equation}
S(J)=\sum_{j_{\text{b}}=0}^{J}\rho(j_{\text{b}}),\qquad S(3j)=1.
\label{eq:normalized_spin_cumulative_sum}
\end{equation}
The figures display $\operatorname{Re}\rho(j_{\text{b}})$ and $\operatorname{Re}S(J)$. The unrestricted and projected amplitudes are each normalized by their own total amplitude. These diagnostics reveal cancellations in the displayed real parts; identifying the associated stationary configurations calls for a complementary critical-point analysis.

\begin{figure*}[tp]
\begin{minipage}{\textwidth}
\centering

\begin{subfigure}[b]{0.48\textwidth}
    \centering
    \includegraphics[width=\textwidth]
    {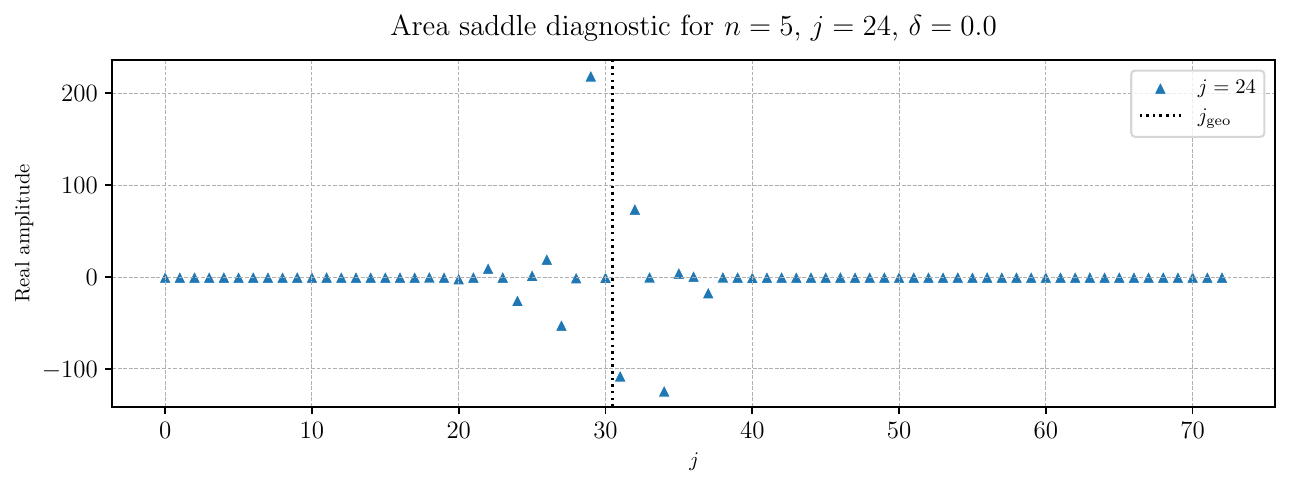}
    \caption{Unrestricted amplitude}
\end{subfigure}
\hfill
\begin{subfigure}[b]{0.48\textwidth}
    \centering
    \includegraphics[width=\textwidth]
    {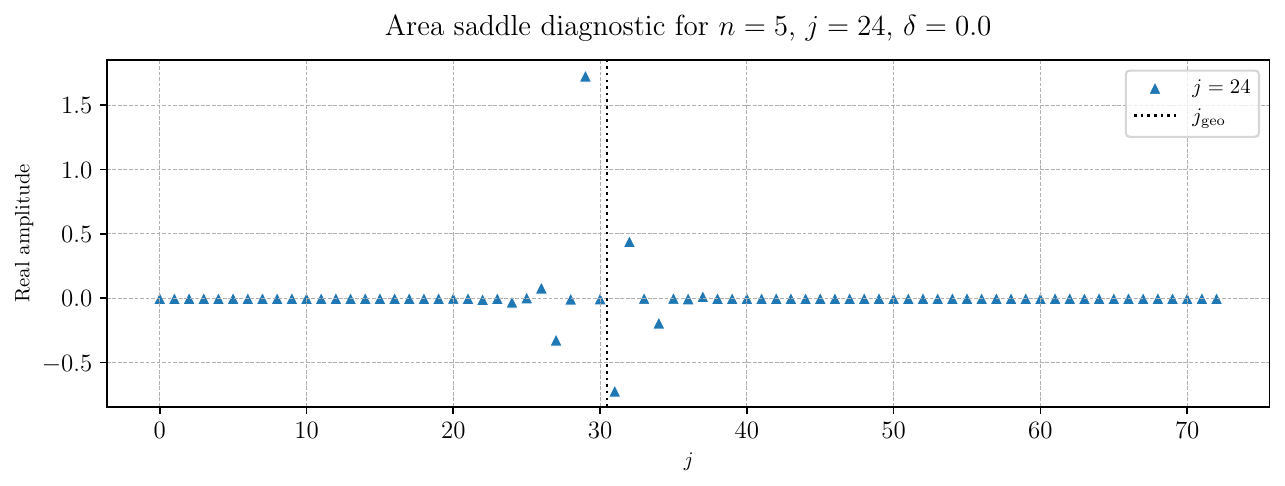}
    \caption{Projected amplitude}
\end{subfigure}

\caption{
Real parts of the normalized spin contributions $\rho(j_{\text{b}})$
for $n=5$, $j=24$, and $\delta=0$. The unrestricted result contains
positive and negative values with magnitudes much larger than unity;
after projection, the displayed values are of order unity.
}
\label{fig:saddle_comparison_n5_j24_d000}
\end{minipage}
\par\medskip
\begin{minipage}{\textwidth}
\setcounter{subfigure}{0}
\centering

\begin{subfigure}[b]{0.48\textwidth}
    \centering
    \includegraphics[width=\textwidth]
    {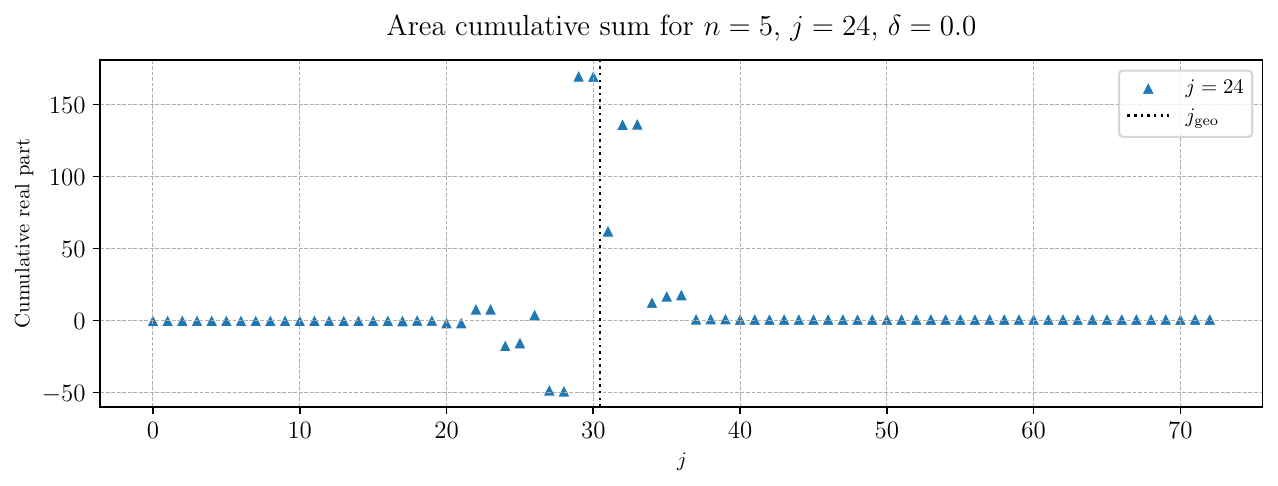}
    \caption{Unrestricted amplitude}
\end{subfigure}
\hfill
\begin{subfigure}[b]{0.48\textwidth}
    \centering
    \includegraphics[width=\textwidth]
    {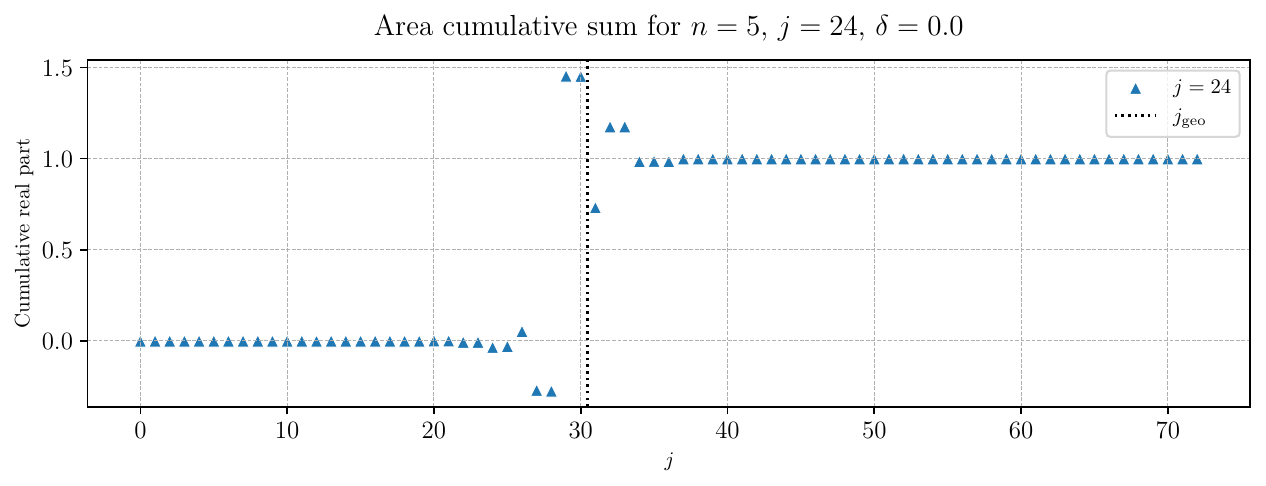}
    \caption{Projected amplitude}
\end{subfigure}

\caption{
Real parts of the normalized cumulative sums $S(J)$ for
$n=5$, $j=24$, and $\delta=0$. The unrestricted sum makes large
excursions before reaching $\operatorname{Re}S(3j)=1$;
the projected sum remains much closer to this endpoint value.
}
\label{fig:cumsum_comparison_n5_j24_d000}
\end{minipage}
\end{figure*}

The spin-contribution plots in \cref{fig:saddle_comparison_n5_j24_d000} show $\operatorname{Re}\rho(j_{\text{b}})$ for $n=5$, $j=24$, and $\delta=0$. In the unrestricted case, several positive and negative terms greatly exceed unity and cancel in the final sum. These terms can lie within a narrow interval around the geometric reference. After projection, the displayed contributions are reduced to order unity.

The cumulative-sum plots in \cref{fig:cumsum_comparison_n5_j24_d000} record the successive sums of the real parts defined in \cref{eq:normalized_spin_cumulative_sum}. A plateau indicates that subsequent terms have little net effect on the displayed sum. For the unrestricted amplitude, excursions of order $10^2$ return to unity through cancellation between positive and negative contributions. Projection substantially reduces these excursions.

Together, these diagnostics associate the negative expectation value and dispersion spike in the unrestricted $n=5$, $\delta=0$ case with large cancellations among normalized spin contributions. The intertwiner restriction substantially changes this cancellation pattern. Competing saddles and vector geometries are possible mechanisms whose role can be tested through critical-point and geometric analysis. For $n=3$, the bulk-spin response to projection is weaker, suggesting a useful comparison between the contributing configurations of the different triangulations.

\Needspace{8\baselineskip}
\section{Discussion} 
\label{sec:ch5:discussion}

This work explores the approach to semiclassical behavior through bulk-spin and intertwiner observables on a family of symmetric $\SU(2)$ BF triangulations. A central finding is that the internal intertwiner remains active even when the boundary spins and coherent geometry are fixed. Its expectation value depends on the deficit angle, spin scale, and triangulation size (\cref{sec:ch5:results:int_raw}), supporting a joint analysis of the bulk-spin and intertwiner sectors.

The unrestricted observables show a close qualitative correspondence. Regions where the bulk-spin ratio approaches unity are generally accompanied by similar behavior in the intertwiner ratio. The clearest departure occurs for $n=5$, $\delta=0$, where the negative bulk-spin expectation value coincides with a shift in the intertwiner ratio and increased dispersions. The flat $n=7$ configuration shows a milder effect, while the $n=9$ bulk-spin ratio remains close to the geometric reference throughout the range considered.

The projector provides a direct way to test the influence of one shape channel centered on the classical reference. For $n=5$, it removes the negative flat-configuration value and moves the bulk-spin ratio toward unity. The projected intertwiner ratios also lie near their targets, with small $D_i$ (\cref{app:numerics:sec:projected_intertwiner_results}), as expected from the prescribed window. For $n=7$, the residual bulk-spin discrepancy is essentially unchanged; for $n=3$, the departure from the geometric value begins at larger deficit angles; and for $n=9$, the ratio remains near unity.

The contribution and cumulative-sum plots clarify this response. For the unrestricted $n=5$, $j=24$, $\delta=0$ amplitude, individual real contributions greatly exceed unity and the cumulative sum makes excursions of order $10^2$. Projection substantially reduces both. This gives a concrete link between the change in bulk-spin moments and the cancellation pattern. The plots characterize interference; determining whether the contributing terms arise within one stationary region or from several configurations requires a critical-point analysis. Suppression of non-Regge contributions, including vector geometries, remains a possible mechanism to investigate.

The exploratory comparison in \cref{app:numerics:sec:exploratory_projector_comparisons} shows that all displayed prescriptions remove the negative $n=5$, $\delta=0$ value, while the quantitative agreement depends on the chosen weight. The finite-window projector provides the reference prescription for the main analysis. A systematic study of window width, its spin scaling, center offsets, and alternative recoupling channels is left to future work and would clarify which features of the response persist across restrictions.

The projector controls one shape channel and uses the classical geometry as input. Establishing a geometric semiclassical sector also involves its conjugate shape variable and shape matching between tetrahedra. Likewise, the amplitude-weighted dispersions can reflect phase cancellations as well as the spread of contributing labels. These distinctions guide the interpretation of the projected moments and the next geometric checks.

The present finite-spin, finite-refinement results provide benchmarks for studying the combined large-spin and refinement regime discussed in the spin-foam literature \cite{Han:2013hga,Han:2017cwc,Han:2021kmi,Asante:2020txa,Asante:2021pxr,Asante:2020sfl}. They highlight the interplay of near-classical bulk observables, intertwiner dynamics, and spin-sum cancellations, particularly around flat configurations. Extending the parameter range together with phase and geometric analysis would connect these observations more directly to the proposed double-scaling limit.

The same strategy can be applied to gravitational spin-foam models. $\SU(2)$ BF theory and the EPRL-FK model share the $\SU(2)$ boundary Hilbert space and intertwiner degrees of freedom, which motivates tracking bulk-spin and intertwiner observables together in EPRL-FK calculations. The projector offers a practical way to test the influence of selected shape channels; whether the specific behavior observed here persists under gravitational dynamics is a question for that comparison.

A natural next step is to repeat this analysis in the Lorentzian EPRL-FK model. This is technically more demanding, both because the vertex amplitude is substantially more expensive to evaluate and because one must construct boundary states that are sufficiently well adapted to the Lorentzian geometry under consideration. The latter point is essential: if the boundary state does not select the intended extrinsic geometry sharply enough, it may obscure precisely the semiclassical effects one wants to test. At the same time, the Lorentzian model offers a possible advantage. Its geometric critical points involve nontrivial $\mathrm{SL}(2,\mathbb{C})$ holonomies and boost data, whereas vector geometries correspond to degenerate configurations with vanishing boosts. This additional structure may provide a cleaner way to distinguish geometric contributions from vector-geometry contributions than is available in the purely $\SU(2)$ BF setting. Whether Lorentzian dynamics actually produce such isolation remains an open problem, but the diagnostics developed here provide a concrete way to test it.

\section*{Acknowledgements}
We thank Pietropaolo Frisoni, Jared Wogan, Sebastian Steinhaus, Rudi Siebert, and Eugenio Bianchi for useful discussions at a preliminary stage of this work. This work was made possible through the support of the WOST, \href{https://withoutspacetime.org}{WithOut SpaceTime project}, supported by Grant ID~\#63683 from the John Templeton Foundation. 
AK is supported by an NSERC grant awarded to Lee Smolin. Research at Perimeter Institute is supported in part by the Government of Canada through the Department of Innovation, Science and Economic Development Canada and by the Province of Ontario through the Ministry of Colleges and Universities. The authors acknowledge the Anishinaabek, Haudenosaunee, L\=unaap\'eewak, Attawandaron, and Neutral peoples, on whose traditional lands Perimeter Institute and the University of Waterloo are located.

\onecolumngrid
\medskip
\twocolumngrid
\appendix
\section{BF Amplitude and Recoupling Structure}
\label{app:bf}

\subsection{Definition of the BF Amplitude}

The SU(2) BF spinfoam amplitude on a given 2--complex is defined in terms of spins and intertwiners assigned to faces and edges of the complex. Each face $f$ is labeled by an SU(2) representation $j_f$, while each edge $e$ carries an intertwiner $i_e$, describing the invariant coupling of the representations associated with the faces incident on that edge.

The vertex amplitudes are associated with 4--simplices of the triangulation and encode the local coupling between the intertwiners of adjacent tetrahedra. The full amplitude is obtained by summing over the spins and intertwiners associated with internal faces and edges, i.e.\ those not fixed by the boundary data, with contributions weighted by the product of vertex amplitudes. In schematic form,
\begin{equation}
\mathcal{A} = \sum_{\{ j_f,\, i_e \}} \prod_v \mathcal{A}_v(j_f, i_e)\,,
\end{equation}
where the sum runs over all internal degrees of freedom. The face weights and the basis-dependent edge weights in \cref{eq:amplitude_definition} are suppressed in this schematic expression.

For a 4-simplex $v$, the boundary consists of five tetrahedra $e_1,\dots,e_5$, and each pair of tetrahedra shares exactly one triangle, giving ten shared triangles $f_1,\dots,f_{10}$ in total. The vertex amplitude $\mathcal{A}_v$ is obtained by contracting the five corresponding 4-valent intertwiners $i_{e_1},\dots,i_{e_5}$ through their ten shared spins $j_{f_1},\dots,j_{f_{10}}$. Since it depends on $10+5=15$ labels, this object is precisely the $\SU(2)$ $\{15j\}$-symbol,
\begin{equation}
\mathcal{A}_v(j_f,i_e) = \{15j\}\bigl(\{j_{f_1},\dots,j_{f_{10}}\};\{i_{e_1},\dots,i_{e_5}\}\bigr)\,,
\label{eq:vertex_15j}
\end{equation}
the $\SU(2)$ analog, one dimension up, of the $\{6j\}$-symbol appearing as the tetrahedron (Ponzano--Regge) vertex amplitude in three-dimensional BF theory, and first considered for four-dimensional $\SU(2)$ BF theory in \cite{Ponzano:1969tw,Ooguri:1992eb}.

Using the recoupling basis described below (\cref{eq:j1_j2,eq:angle_operator,eq:Casimir}), each of the five 4-valent intertwiners $i_{e_a}$ can be further split into a pair of 3-valent vertices joined by an internal edge carrying the recoupling spin $i_{e_a}$ itself, turning the boundary graph of the 4-simplex into a fully trivalent $\SU(2)$ spin network with ten nodes and fifteen edges. This is precisely the pentagon graph associated with the irreducible $\{15j\}$-symbol of the first kind \cite{Varshalovich:1988vb,Yutsis:1962wm}, which reduces to a single sum over one auxiliary spin $x$ of a product of five $\{6j\}$-symbols,
\begin{widetext}
\begin{equation}
\{15j\}\bigl(\{j_a\},\{l_a\},\{k_a\}\bigr) = (-1)^{\sum_{a=1}^{5} (j_a+l_a+k_a)} \sum_{x} (2x+1) \prod_{a=1}^{5} \left\lbrace\begin{array}{ccc} j_a & k_a & x \\ k_{a+1} & j_{a+1} & l_a \end{array}\right\rbrace\,,
\label{eq:15j_via_6j}
\end{equation}
\end{widetext}
with the indices taken cyclically ($a=1,\dots,5$ mod $5$) and the identification of the fifteen physical spins $\{j_f,i_e\}$ with the array entries $\{j_a\},\{l_a\},\{k_a\}$ fixed once a choice of recoupling channels for the five intertwiners is made \cite{Dona:2022dxs}. This reduction underlies the graphical evaluation of the amplitude described in \cref{app:graphical_rules}.

A lower-dimensional visualization uses triangles glued around a vertex in two dimensions. If the sum of their internal angles at that vertex is $2\pi$, the neighborhood is locally flat. An angular deficit or excess represents curvature concentrated at the vertex.




\subsection{Recoupling Basis and Intertwiners}

Intertwiners provide the building blocks for constructing invariant amplitudes in SU(2) BF theory and encode the coupling of representations associated with adjacent faces.

Given spins $j_1, \dots, j_n$ incident at an edge, an intertwiner is an invariant tensor in the space
\begin{equation}
i \in \mathrm{Inv}\left(V^{j_1} \otimes \cdots \otimes V^{j_n}\right),
\end{equation}
where $V^j$ denotes the representation space of spin $j$.

In the case of a 4--simplex, the relevant intertwiners are 4--valent. A convenient basis is provided by a recoupling scheme, in which representations are combined pairwise according to SU(2) Clebsch--Gordan rules. In this basis, the intertwiner is labeled by an intermediate spin corresponding to the chosen coupling channel.

We write $d_i=2i+1$ and use the graphical basis $\ket{i}$ obtained by contracting two normalized $3j$-symbols through the invariant tensor,
\begin{equation}
\begin{aligned}
\iota^{(i)}_{m_1m_2m_3m_4}
&=\sum_m(-1)^{i-m}\\
&\quad\times
\begin{pmatrix}j_1&j_2&i\\m_1&m_2&m\end{pmatrix}
\begin{pmatrix}i&j_3&j_4\\-m&m_3&m_4\end{pmatrix}.
\end{aligned}
\label{eq:graphical_intertwiner_tensor}
\end{equation}
Its normalization and the associated resolution of the identity are \cite{Makinen:2019rou}
\begin{equation}
\braket{i}{i'}=\frac{\delta_{ii'}}{d_i},\qquad
\sum_i d_i\ket{i}\bra{i}=\mathds{1}_{\mathrm{Inv}}.
\label{eq:intertwiner_basis_norm}
\end{equation}
The corresponding orthonormal basis is
\begin{equation}
\ket{i}_{\mathrm{N}}=\sqrt{d_i}\ket{i},\qquad
{}_{\mathrm{N}}\!\braket{i}{i'}_{\mathrm{N}}=\delta_{ii'}.
\label{eq:orthonormal_intertwiner_basis}
\end{equation}
The $\{15j\}$-symbol in \cref{eq:vertex_15j,eq:15j_via_6j} is the contraction of five graphical basis tensors. For the same choice of coupling channels on both sides, replacing them with orthonormal intertwiners gives $\widehat{\mathcal{A}}_v=(\prod_{e\subset v}\sqrt{d_{i_e}})\mathcal{A}_v$. In \cref{eq:coherent_vertex_normalized}, $\widehat{\mathcal{A}}_v$ is expressed in the gluing channels after the transformation in \cref{eq:normalized_recoupling}. This conversion also accounts for the edge weights: a pair of factors $\sqrt{d_i}$ at the two ends of an internal edge supplies the graphical-basis weight $d_i$. The numerical vertex, recoupling matrix, and coherent coefficients all use this orthonormal convention.

A particularly useful property of the recoupling basis is that it diagonalizes the operator $\vec{J}_1 \cdot \vec{J}_2$, where $\vec{J}_1$ and $\vec{J}_2$ are the SU(2) generators associated with two of the faces meeting at the node. In the recoupling basis labeled by the intermediate spin $i$, one has the exact eigenvalue relation
\begin{equation}
\vec{J}_1 \cdot \vec{J}_2
=
\frac{1}{2}
\left[
i(i+1)
-
j_1(j_1+1)
-
j_2(j_2+1)
\right].
\label{eq:j1_j2}
\end{equation}

The geometric interpretation of this relation follows from the external dihedral angle operator
\begin{equation}
\cos\theta_{\rm ext}
=
\frac{\vec{J}_1 \cdot \vec{J}_2}
{\sqrt{J_1^2 J_2^2}},
\label{eq:angle_operator}
\end{equation}
together with
\begin{equation}
J_a^2=j_a(j_a+1).
\label{eq:Casimir}
\end{equation}
The external angle $\theta_{\rm ext}$ is the angle between the outward face normals associated with the two representations. Since the recoupling basis diagonalizes $\vec{J}_1 \cdot \vec{J}_2$, different intertwiner labels correspond to different angle eigenvalues.

Combining \cref{eq:j1_j2,eq:angle_operator,eq:Casimir} gives
\begin{equation}
\begin{aligned}
i(i+1)&=j_1(j_1+1)+j_2(j_2+1)\\
&\quad+2\sqrt{j_1(j_1+1)\,j_2(j_2+1)}
\cos\theta_{\rm ext},
\end{aligned}
\end{equation}
which provides the exact relation between the intertwiner label and the corresponding angle.

For the symmetric configurations considered in this work,
\begin{equation}
j_1=j_2=j,
\end{equation}
and the relation simplifies to
\begin{equation}
\begin{aligned}
i(i+1)&=2j(j+1)+2j(j+1)\cos\theta_{\rm ext}\\
&=2j(j+1)\bigl(1+\cos\theta_{\rm ext}\bigr).
\end{aligned}
\end{equation}

From a geometric perspective, intertwiners encode the shape degrees of freedom of a tetrahedron. In particular, one can relate the intertwiner to the dihedral angle between the corresponding faces.

One must distinguish between the external and internal dihedral angles. The external dihedral angle $\theta_{\text{ext}}$ is the angle between outward normals to adjacent faces of a tetrahedron, while the internal dihedral angle is its supplement,
\begin{equation}
\theta_{\text{int}} = \pi - \theta_{\text{ext}}\,,
\qquad
\cos\theta_{\text{int}} = -\cos\theta_{\text{ext}}\,.
\end{equation}

In this work, we adopt the convention in which $\theta$ denotes the external dihedral angle. With this choice, the intertwiner is related to the geometry by
\begin{equation}
i(i+1) = 2 j(j+1)\bigl(1 + \cos\theta\bigr)\,.
\label{eq:intertwiner_external}
\end{equation}

The recoupling basis plays a central role in the numerical implementation, as it allows one to express both vertex amplitudes and intertwiner variables in a form suitable for summation and projection.

\subsection{Graphical Representation and Rules}
\label{app:graphical_rules}

Graphical rules provide a convenient representation of SU(2) invariant structures and are widely used to simplify recoupling relations in spinfoam amplitudes. In this framework, representations are associated with edges, intertwiners with vertices, and contractions with the gluing of graphical elements.

In the present work, these graphical methods play a direct computational role. While vertex amplitudes are evaluated numerically using \texttt{sl2cfoam-next}, the graphical decomposition determines the recoupling structure, phase relations, and contraction patterns entering the calculation. The graphical rules summarized below therefore serve both as a tool for deriving the factors appearing in the amplitudes and as a guide for constructing the quantities supplied to the numerical implementation.

The graphical rules summarized here are largely based on the presentation of \cite{Makinen:2019rou}, which provides a clear and pedagogical introduction to the graphical calculus for SU(2) representations. The figures collected below should therefore be viewed primarily as a reference sheet of the graphical identities used throughout the derivations, rather than as a self-contained introduction to the graphical calculus.

The identities below are organized by their role in the graphical calculus. The elementary graphical objects, including Kronecker deltas, loops, orientation tensors, and their composition and inversion properties, are shown in \cref{fig:edge,fig:loop,fig:edge_directed,fig:edge_inversion,fig:edge_composition,fig:edge_inverted_composition}. Trivalent vertices and their symmetry properties are summarized in \cref{fig:vertex_trivalent,fig:edge_trivalent_permutation,fig:edge_trivalent_directed}. \cref{fig:bubble} and \cref{fig:normalization} give diagrammatic simplifications and normalization conventions. Finally, the graphical representation of group elements, their inversion properties, recoupling expansions, and the corresponding Haar integration identities are collected in \cref{fig:group_element,fig:group_element_inverted,fig:group_element_double_integral,fig:group_element_double,fig:group_element_trivalent_vertex,fig:group_element_quadrapule_integral,fig:group_element_triple_integral}. Together, these relations provide the graphical machinery required to manipulate spin-network diagrams and to determine the multiplicative factors, such as loop, bubble, and theta-network evaluations (see \cref{fig:theta_network}), that arise during the recoupling procedure.
\onecolumngrid
\refpair{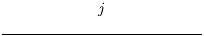}{$\delta_{mn}$}{fig:edge}%
        {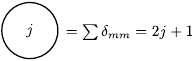}{A loop.}{fig:loop}
\refpair{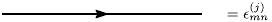}{$\epsilon^{(j)}_{m n}$ in irrep $j$.}{fig:edge_directed}%
        {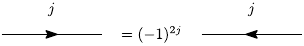}{Inversion of an irrep $j$ line.}{fig:edge_inversion}
\refpair{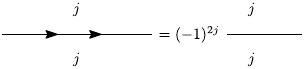}{Composition of two equally oriented $\epsilon$'s.}{fig:edge_composition}%
        {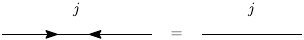}{Composition of two oppositely oriented $\epsilon$'s.}{fig:edge_inverted_composition}
\refpair{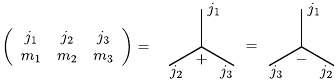}{$3j$-symbol. The $+$ denotes clockwise and the $-$ counterclockwise rotation.}{fig:vertex_trivalent}%
        {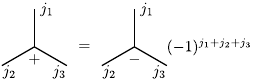}{Permutation of two $\delta$'s in a trivalent vertex.}{fig:edge_trivalent_permutation}
\refpair{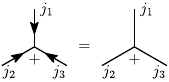}{Trivalent vertex of three $\epsilon$'s.}{fig:edge_trivalent_directed}%
        {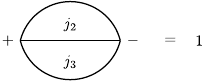}{Normalization.}{fig:normalization}
\refsingle[1]{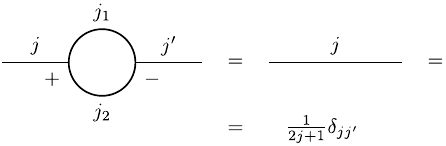}{Removal of a bubble.}{fig:bubble}
\refpair{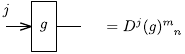}{Group element.}{fig:group_element}%
        {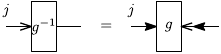}{Inverse of a group element.}{fig:group_element_inverted}
\refsingle{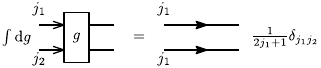}{Integral of the contraction of two representations (Wigner matrices) of a group element in the Clebsch–Gordan series, i.e., the orthogonality theorem of two Wigner matrices.}{fig:group_element_double_integral}
\refsingle{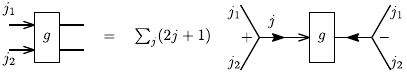}{Contraction of two representations (Wigner matrices) of a group element in the Clebsch–Gordan series.}{fig:group_element_double}
\refsingle{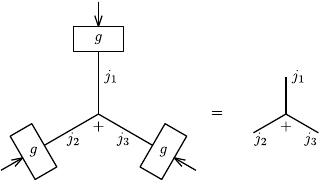}{Group element trivalent vertex to simplex trivalent vertex.}{fig:group_element_trivalent_vertex}
\refsingle{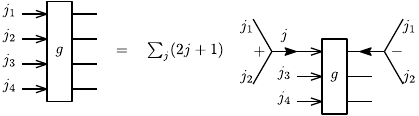}{Contraction of four representations (Wigner matrices) of a group element.}{fig:group_element_quadrapule_integral}
\refsingle{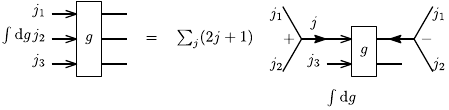}{Haar integral of a contraction of three representations (Wigner matrices) of a group element.}{fig:group_element_triple_integral}
\refsingle[1]{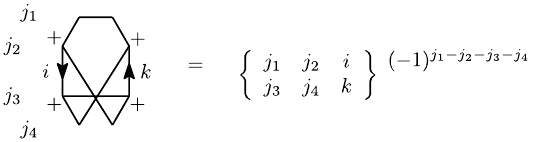}{A theta network arising during the recoupling procedure. Its evaluation contributes a multiplicative factor determined by the graphical identities summarized in this appendix.}{fig:theta_network}
\twocolumngrid

\section{Classical Geometry and Regge Structure}
\label{app:geometry}

\subsection{Geometry of a 4--Simplex and Its Boundary}

A 4--simplex is the fundamental building block of the triangulations considered in this work. It consists of five vertices, ten edges, ten triangular faces, and five tetrahedral boundary components.

The geometry of a 4--simplex can be specified by assigning geometric data such as edge lengths, or equivalently, areas and dihedral angles subject to consistency conditions. Each triangular face is associated with an area variable, while tetrahedra correspond to collections of four triangles and encode shape information.

The geometry of the 4--simplex is characterized by the dihedral angles between its tetrahedral faces. These angles encode how the tetrahedra are glued together and play a central role in both classical and quantum constructions.

\subsection{Gluing of Simplices and Deficit Angle}

Extended triangulations are constructed by gluing multiple 4--simplices along shared tetrahedra, forming a simplicial complex.

In such constructions, curvature is not associated with the interior of simplices, which are flat, but is instead localized on lower-dimensional subsimplices, in particular on triangles.

When several 4--simplices meet at a common triangle, each simplex contributes a 4D dihedral angle dual to that triangle. The total angle around the triangle equals the sum of the 4D dihedral angles from all simplices that share it.

The deficit angle $\delta$ is defined as
\begin{equation}
\delta = 2\pi - \sum_i \theta_i\,,
\end{equation}
where $\theta_i$ is the internal four-dimensional dihedral angle at the triangle in simplex $i$, and the sum runs over all simplices incident on that triangle.

A vanishing deficit angle corresponds to a flat configuration, while a nonzero value indicates the presence of curvature localized at the triangle.

In the configurations considered in this work, we construct a family of triangulations $\Delta_n$ by gluing $n$ identical 4--simplices along a common triangle. In this setting, the deficit angle depends on the number of simplices and the geometry of each building block, and serves as the parameter controlling the configuration's curvature.

\subsection{Symmetric Configurations Used in This Work}
\label{app:geometry:sec:symmetric}

The configurations considered in this work are based on a symmetry-reduced 4--simplex, in which the ten edge lengths are grouped into a small number of equivalence classes. This reduction preserves the configuration's symmetry while significantly reducing the number of independent geometric variables.

More specifically, we implement the symmetry reduction by introducing three edge-length parameters. The three edges belonging to the triangle shared by all simplices are assigned a common length $x$. The six edges connecting the vertices of this triangle to the two remaining vertices are assigned a common length $a$. The remaining edge, connecting the two vertices not in the shared triangle, is assigned length $b$. In the reconstruction notebook associated with the repository \cite{double_limit_repository}, the variables \texttt{xs}, \texttt{as}, and \texttt{bs} denote the squared lengths $x^2$, $a^2$, and $b^2$, respectively. This parametrization distinguishes the shared triangle from the boundary geometry already at the level of the edge-length data and reduces the simplex geometry to three independent variables.

The symmetry relates the three external tetrahedra to each other and likewise the two internal tetrahedra. These form two distinct classes: each external tetrahedron has four boundary triangles, whereas each internal tetrahedron contains the shared triangle and three boundary triangles. The remaining freedom can be interpreted in terms of an overall scale, fixing the size of the configuration, together with shape parameters determining the relative values of the three edge-length classes $(x,a,b)$.

An additional constraint is imposed by requiring all boundary triangles to have equal area. In general, different triangles of a 4--simplex would have different areas due to the independent assignment of edge lengths. In the present construction, relations among the edge lengths are imposed so that all boundary triangle areas coincide. A reference area is then fixed in order to remove the overall rescaling freedom of the geometry.

The equal-area and deficit-angle conditions determine the geometry only
up to a common rescaling of all edge lengths. Under such a rescaling,
all triangle areas are multiplied by the same factor, while area ratios
and dihedral angles remain unchanged. Since there is only a single global
scaling degree of freedom, it is sufficient to fix the common area of
the boundary triangles. In the reconstruction procedure, this common
boundary area is set equal to unity.

We prescribe a deficit angle $\delta$ which uniquely determines the corresponding geometry within our family of symmetric triangulations. For a configuration consisting of $n$ identical simplices glued around a common triangle, this fixes the corresponding internal 4--dimensional dihedral angle $\alpha$ according to 
\begin{equation}
    \alpha=\frac{2\pi-\delta}{n}.
\end{equation}

The geometry is therefore specified by the symmetry assumptions, the equal-area conditions, and the prescribed value of the deficit angle.

Starting from the reduced edge-length parametrization, the simplices are reconstructed by seeking a consistent embedding in four-dimensional Euclidean space. This amounts to assigning coordinates to the vertices such that the prescribed edge lengths, areas, and geometric relations are simultaneously satisfied. The corresponding areas, volumes, and dihedral angles are then computed and used to construct a system of nonlinear consistency conditions for the remaining geometric parameters. This system is solved numerically. In general, multiple solution branches may be present. The branch employed in this work is selected according to the geometric reconstruction procedure used in previous studies and yields a consistent Euclidean simplicial geometry \cite{Dona:2019jab,Dona:2020tvv}. In practice, this corresponds to the solution whose reconstructed embedding is characterized by predominantly positive coordinate values. The dependence of the results on the choice of branch is not investigated here.

Once the geometry is reconstructed, the area of the shared triangle is
computed from the resulting configuration. Because the common boundary
area has been fixed to unity, this quantity may be interpreted as the
shared-triangle area measured in units of the boundary area. Denoting
this scale-independent quantity by $A_{\text{shared}}$, the
corresponding classical bulk-spin value for a boundary spin $j$, which sets the scale, is
defined by
\begin{equation}
j_{\text{classical}}
=
j\,A_{\text{shared}}.
\end{equation}
This reference uses the leading large-spin identification of area with $j$. The reconstructed geometry also determines the tetrahedral dihedral angles and the geometry of the external tetrahedra. These data are then used as input for the coherent boundary states entering the spinfoam amplitude and provide the classical reference configuration used throughout this work.

\section{Intertwiner Sector and Projector}
\label{app:intertwiner}

\subsection{Intertwiners in the Recoupling Basis}
\label{app:intertwiner:sec:recoupling}

A bulk tetrahedron carries a four-valent $\SU(2)$ intertwiner with face spins $(j,j,j,j_{\text{b}})$. A convenient basis is provided by a recoupling scheme, in which the representations are combined pairwise according to Clebsch--Gordan rules. In this basis, the intertwiner space is spanned by states labeled by an intermediate spin $i_{\text{b}}$. For general face spins $j_1,j_2,j_3,j_4$, the allowed values of $i_{\text{b}}$ are determined by the triangle inequalities,
\begin{equation}
\begin{split}
i_{\text{b}} \in &\{|j_1-j_2|,\dots,j_1+j_2\}\\
\cap &\{|j_3-j_4|,\dots,j_3+j_4\}.
\end{split}
\end{equation}

Each range advances in unit steps and must satisfy the usual parity conditions: $j_1+j_2+i_{\text{b}}$ and $j_3+j_4+i_{\text{b}}$ are integers.

In this representation, the intertwiner appears as a discrete label of the recoupling basis on which the amplitude depends. Although the full amplitude involves a summation over such labels, in this work the dependence on the intertwiner is evaluated explicitly before performing this summation, allowing access to the corresponding amplitude-weighted moments.

\subsection{Geometric Interpretation of the Intertwiner}

In a geometric interpretation, intertwiners are associated with the tetrahedra of the triangulation and encode their shape degrees of freedom.

A tetrahedron is characterized by the areas of its faces together with additional data specifying how these faces are glued together. While the face areas determine the magnitudes of the corresponding geometric quantities, they do not uniquely fix the geometry of the tetrahedron.

This can be contrasted with the usual formulation of Regge calculus in terms of edge lengths, where the geometry of a simplex is uniquely determined by the assignment of its edge lengths, up to a finite set of geometrically equivalent embeddings related by orientation or reflection. When working with areas instead, this is no longer the case: specifying the areas of the faces does not uniquely determine the geometry of the tetrahedron. Additional information, such as dihedral angles or equivalently intertwiner variables, is required to fully reconstruct the shape.

The remaining degrees of freedom are encoded in the relative orientation of the faces, which can be described in terms of dihedral angles between them. Geometrically, the dihedral angle measures the relative orientation of two faces sharing a common edge, and can be understood as the angle between the planes defined by the corresponding triangles.

At fixed face areas, a nondegenerate tetrahedron has a two-dimensional classical shape space \cite{Freidel:2010aa}. The recoupling label $i$ diagonalizes one pairwise area-vector product, or equivalently one face-angle combination. A general intertwiner state encodes quantum shape information in its basis coefficients; full geometric reconstruction also involves the conjugate shape variable and matching conditions between tetrahedra.

In the present work, this correspondence is made explicit by relating the intertwiner to dihedral angles computed from the classical geometry described in \cref{app:geometry}.

\subsection{Classical Value of the Intertwiner}

The classical value of the intertwiner is determined by its relation to the geometry of the tetrahedron, in particular through the dihedral angle between its faces.

As discussed in \cref{app:bf}, this relation follows from the addition of angular momenta and can be expressed in terms of the angle between the corresponding geometric quantities. For the case of equal spins $j_1 = j_2 = j$, \cref{eq:intertwiner_external} gives, in terms of the external dihedral angle,
\begin{equation}
i(i+1) = 2 j(j+1)\bigl(1 + \cos\theta_{\text{ext}} \bigr),
\end{equation}
which is the convention adopted in this work.

We define $i_{\text{classical}}$ as the nonnegative solution of this equation, using the external angle between the two boundary faces paired in the chosen recoupling channel. The angle is computed from the geometry reconstructed in \cref{app:geometry}.

\subsection{Definition of the Projector}

The motivation for restricting the intertwiner sum, and the resulting projected amplitude are discussed in \cref{ch5:sec:numerics:subsec:restricting}. For reference, the finite-window projector used throughout the main analysis restricts the intertwiner sum to
\begin{equation}
i_{\text{b}} \in [y-1,\, y+2]\,, \qquad y = \lfloor i_{\text{classical}} \rfloor\,,
\end{equation}
where $i_{\text{classical}}$ is the classical intertwiner value obtained from the reconstructed geometry (\cref{app:geometry}). The window is intersected with the admissible labels at each bulk spin. Its center supplies a classical input and at fixed geometry its fixed span restricts the relative label support increasingly as $j$ grows. It constrains one shape channel and replaces the full intertwiner identity with a projector, thereby modifying the BF amplitude. Alternative projector prescriptions and their comparison against this one are given in \cref{app:numerics:sec:exploratory_projector_comparisons}.

In the orthonormal gluing basis of \cref{eq:coherent_vertex_normalized}, let $W(j_{\text{b}})$ denote this admissible window and let $\chi_W(i)$ be its indicator. The diagonal projector has entries $P(j_{\text{b}})_{ab}=\delta_{ab}\chi_W(a)$. Restricting every bulk intertwiner label gives
\begin{equation}
M_P(j_{\text{b}})=P(j_{\text{b}})M(j_{\text{b}})P(j_{\text{b}}),
\label{eq:projected_vertex_matrix}
\end{equation}
and the corresponding partial amplitude is
\begin{equation}
a_P(j_{\text{b}})=d_{j_{\text{b}}}\operatorname{Tr}\!\left[M_P(j_{\text{b}})^n\right].
\label{eq:projected_partial_amplitude}
\end{equation}
Here the channel change in \cref{eq:normalized_recoupling} precedes projection, so $P$ acts on the labels used for gluing. Its idempotence ensures that each label is restricted once in the cyclic contraction.

\section{Numerical Implementation Details}
\label{app:numerics}

This appendix describes the numerical implementation of the quantities defined in \cref{app:bf,app:intertwiner}.

\subsection{Definition of Observables}

The bulk-spin expectation value and its relative dispersion are defined in \cref{sec:ch5:config}. For observables involving the internal intertwiners, retain the full vector $\boldsymbol{i}_{\text{b}}=(i_{\text{b}_1},\ldots,i_{\text{b}_n})$. Let $\mathcal{W}(j_{\text{b}},\boldsymbol{i}_{\text{b}})$ be the summand of \cref{eq:amplitude_deltan}, including its face and edge weights and its boundary-contracted vertex factors. A general observable $O(j_{\text{b}},\boldsymbol{i}_{\text{b}})$ has expectation value
\begin{equation}
\langle O \rangle =
\frac{\sum_{j_{\text{b}},\,\boldsymbol{i}_{\text{b}}}
O(j_{\text{b}},\boldsymbol{i}_{\text{b}})\,\mathcal{W}(j_{\text{b}},\boldsymbol{i}_{\text{b}})}
{\sum_{j_{\text{b}},\,\boldsymbol{i}_{\text{b}}}
\mathcal{W}(j_{\text{b}},\boldsymbol{i}_{\text{b}})}\,,
\label{eq:general_observable}
\end{equation}
with a nonzero denominator. Summing $\mathcal{W}$ over all entries of $\boldsymbol{i}_{\text{b}}$ at fixed $j_{\text{b}}$ recovers $a(j_{\text{b}})$ of \cref{eq:partial_amplitude}. Taking $O$ to depend only on $j_{\text{b}}$ therefore gives the bulk-spin observables. For the finite-window restriction, multiply the summand by $\prod_{e=1}^{n}\chi_W(i_{\text{b}_e})$.

The intertwiner observables plotted in this work refer to one tagged internal tetrahedron: we write $i_{\text{b}}\equiv i_{\text{b}_1}$ and sum independently over all other labels. Cyclic symmetry makes its moments independent of which tetrahedron is tagged. In the gluing basis, define $I(j_{\text{b}})_{ab}=a\delta_{ab}$, where $a$ is the spin label rather than its array position. The moments are then
\begin{equation}
\langle i_{\text{b}}^p\rangle
=\frac{\sum_{j_{\text{b}}}d_{j_{\text{b}}}
\operatorname{Tr}\!\left[I(j_{\text{b}})^pM(j_{\text{b}})^n\right]}
{\sum_{j_{\text{b}}}a(j_{\text{b}})},\qquad p=1,2.
\label{eq:intertwiner_moments}
\end{equation}
The insertion acts on the recoupled gluing label. For projected moments, replace $M$ by $M_P$ and $a$ by $a_P$ in this expression.

\subsection{Evaluation of Expectation Values}
\label{app:numerics:sec:expvals}

All expectation values are evaluated numerically for each fixed value of the deficit angle and triangulation size, by direct summation over the stored, discretized amplitude.

It is worth emphasizing a structural difference between the treatment of the bulk spin and the intertwiner variables in the numerical implementation. In the spinfoam amplitude, one sums over all internal spins and intertwiners. After fixing the boundary spins and boundary intertwiners, the remaining internal degrees of freedom consist of the bulk spin $j_{\text b}$ and the internal intertwiner variables.

In the implementation used in this work, based on \texttt{sl2cfoam-next}, the vertex amplitudes are provided as multidimensional arrays with the intertwiner dependence represented explicitly as tensor indices. A separate vertex amplitude is computed for each value of the bulk spin $j_{\text b}$, and the associated array indices correspond to the intertwiner degrees of freedom of the tetrahedra of the simplex. After contraction with the boundary states, the remaining internal intertwiner indices are summed over through a trace operation. The resulting traced amplitude is then stored only as a function of $j_{\text b}$.

Consequently, the dependence of the amplitude on the bulk spin is directly accessible. Expectation values of polynomial functions of $j_{\text b}$ can therefore be computed straightforwardly as weighted sums of the form
\begin{equation}
\sum_{j_{\text b}} j_{\text b}^{\,p}\, a(j_{\text b}),
\end{equation}
where $p$ is a nonnegative integer, without the need for further reconstruction from the stored data.

By contrast, the intertwiner variables are not retained explicitly after the trace is performed. Any quantity requiring direct access to the intertwiner dependence, including intertwiner moments and the implementation of intertwiner projectors, must therefore be evaluated prior to the contraction and trace. Derived quantities, such as dispersions, can subsequently be reconstructed from the corresponding moments.

\subsection{Relative Dispersion}
\label{app:numerics:sec:dispersion}

The relative dispersion of the bulk spin, $D_j = \sqrt{\left|\langle j_{\text{b}}^2\rangle/\langle j_{\text{b}}\rangle^2 - 1\right|}$, is defined and interpreted in \cref{sec:ch5:config}; the intertwiner relative dispersion $D_i$ used in \cref{sec:ch5:results:int_raw} is defined analogously, with $j_{\text{b}}\to i_{\text{b}}$. Both are evaluated from the stored, discretized amplitude as functions of the deficit angle and number of simplices. Their denominators must be nonzero, and their amplitude-weighted interpretation follows the discussion in \cref{sec:ch5:config}.

\subsection{Spin-Contribution and Cumulative-Sum Diagnostics}

The spin-contribution and cumulative-sum diagnostics are defined in \cref{eq:normalized_spin_contribution,eq:normalized_spin_cumulative_sum}. The plots show $\operatorname{Re}[a(j_{\text{b}})/\mathcal{A}_{\Delta_n}]$ and the real part of its cumulative sum in ascending bulk-spin order. Each unrestricted or projected result is normalized by its own nonzero total amplitude. Large positive and negative values indicate cancellations among the displayed real contributions.

A cumulative-sum plateau indicates that the remaining terms have little net effect on the displayed sum. Large excursions reflect cancellations, which can occur within one stationary region or between several critical configurations. Their geometric interpretation can be explored through a complementary phase and critical-point analysis.

\cref{sec:ch5:results:saddle} discusses detailed examples and their physical interpretation.

\subsection{Numerical Evaluation Strategy}

We evaluate the amplitude and associated expectation values by explicitly summing over discrete values of the bulk spin and, when applicable, the intertwiner variable.

The summation over the bulk spin $j_{\text{b}}$ runs over a finite range determined by the boundary data. In practice, the amplitude is computed for all values of $j_{\text{b}}$ in a predefined interval, typically starting from $j_{\text{b}} = 0$ up to a maximal value fixed by the chosen boundary spins, which in the present configurations corresponds to $j_{\text{b}} \leq 3j$.

The origin of this bound follows directly from SU(2) recoupling theory. In the symmetric configurations considered here, the bulk spin is obtained by coupling three spins of magnitude $j$. This coupling can be performed iteratively. First, two spins are combined,
\begin{equation}
    j \otimes j \rightarrow k,
\end{equation}
with the intermediate spin taking values
\begin{equation}
    k \in \{0,\dots,2j\}.
\end{equation} 
The resulting spin is then coupled to a third spin $j$. The largest allowed value of the bulk spin is obtained when the intermediate spin reaches its maximal value, $k = 2j$, giving
\begin{equation} 
    j_{\text{b},\text{max}} = 2j + j = 3j.
\end{equation}
Consequently, the range
\begin{equation}
    j_{\text{b}} \in \{0,\dots,3j\}
\end{equation} 
follows exactly from SU(2) recoupling and is not a numerical truncation.





\subsection{Exploratory Projector Comparisons} \label{app:numerics:sec:exploratory_projector_comparisons}

We explored several intertwiner restrictions to assess the results' sensitivity to the choice of intertwiner restriction. This comparison is an exploratory consistency check rather than an exhaustive survey of possible projector constructions. Only prescriptions that produced at least a partial improvement of the observed discrepancy are included. Additional projector choices were tested but are omitted for brevity, and the set considered here should not be regarded as exhaustive. In addition to the unrestricted amplitude ("full") and the finite-window ("hard") projector explained in \cref{ch5:sec:numerics:subsec:restricting}, we consider a cosine-squared weight ("cos2") and a Gaussian weight ("gaussian"), defined respectively by
\begin{equation}
P_{\mathrm{cos2}}(i)
=
\begin{cases}
\cos^2\!\left(
\frac{\pi (i-i_{\mathrm{classical}})}
{2\sigma}
\right),
&
|i-i_{\mathrm{classical}}|\le \sigma,
\\[1ex]
0,
&
\text{otherwise},
\end{cases}
\end{equation}
\begin{equation}
P_{\mathrm{gaussian}}(i)
=
\exp\!\left[
-\frac{(i-i_{\mathrm{classical}})^2}
{2\sigma^2}
\right],
\end{equation}
and
\begin{equation}
P_{\mathrm{hard}}(i)
=
\begin{cases}
1,
&
i\in
[\lfloor i_{\mathrm{classical}}\rfloor-1,
 \lfloor i_{\mathrm{classical}}\rfloor+2],
\\[1ex]
0,
&
\text{otherwise},
\end{cases}
\end{equation}

The parameter $\sigma$ controls the width of the cosine-squared and Gaussian weights around the classical intertwiner value. Only the finite-window prescription is an idempotent projector; the smooth prescriptions act as weights. We pay particular attention to the case $j=24$ and $\delta=0$, since the unrestricted amplitude shows its most pronounced deviations in this regime. To illustrate the dependence on the projector prescription, we compare the normalized area expectation value
\begin{equation}
\frac{\operatorname{Re}\langle j_{\mathrm{b}} \rangle}{j_{\mathrm{classical}}}
\end{equation}
for the cases $n=3$ and $n=5$ at fixed boundary spin $j=24$ and deficit angle $\delta=0$. These configurations provide a useful benchmark: the $n=3$ triangulation already exhibits behavior close to the geometric prediction, whereas the $n=5$ triangulation displays the most pronounced pathology observed in the unrestricted amplitude. The latter therefore provides a particularly sensitive test of the effect of restricting the intertwiner sector.

\begin{table}[tp]
\centering

\begin{tabular}{lcc}
\toprule
Restriction & $n=3$ & $n=5$ \\
\midrule
full & 1.011421 & -18.831057 \\
\midrule
cos2 ($\sigma=1.0$) & 1.041955 & 1.042103 \\

gaussian ($\sigma=0.5$) & 1.046484 & 0.284836 \\

hard (finite window) & 1.039867 & 0.939331 \\
\bottomrule
\end{tabular}
\caption{Comparison of the normalized bulk-spin expectation value $\operatorname{Re}\langle j_{\text{b}}\rangle/j_{\mathrm{classical}}$ for several internal-intertwiner restrictions. Values correspond to boundary spin $j=24$ and deficit angle $\delta=0.00$. The reported quantity is identical to that used in the area-average analysis.}
\label{tab:projector_comparison}
\end{table}

\begin{figure*}[tp]
\centering

\begin{subfigure}[b]{0.48\textwidth}
    \centering
    \includegraphics[width=\textwidth]
    {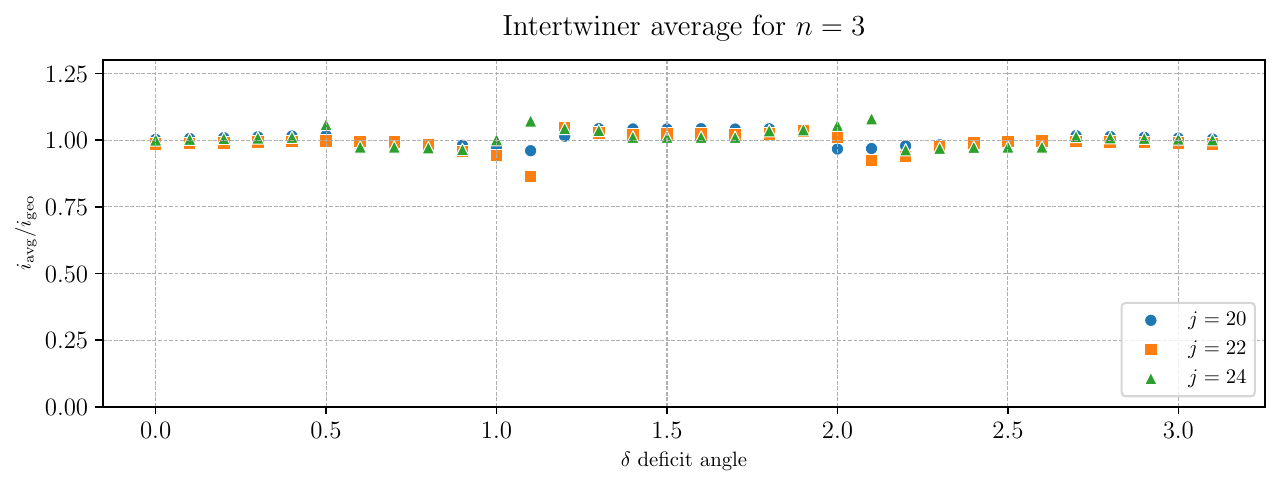}
    \caption{$n=3$}
\end{subfigure}
\hfill
\begin{subfigure}[b]{0.48\textwidth}
    \centering
    \includegraphics[width=\textwidth]
    {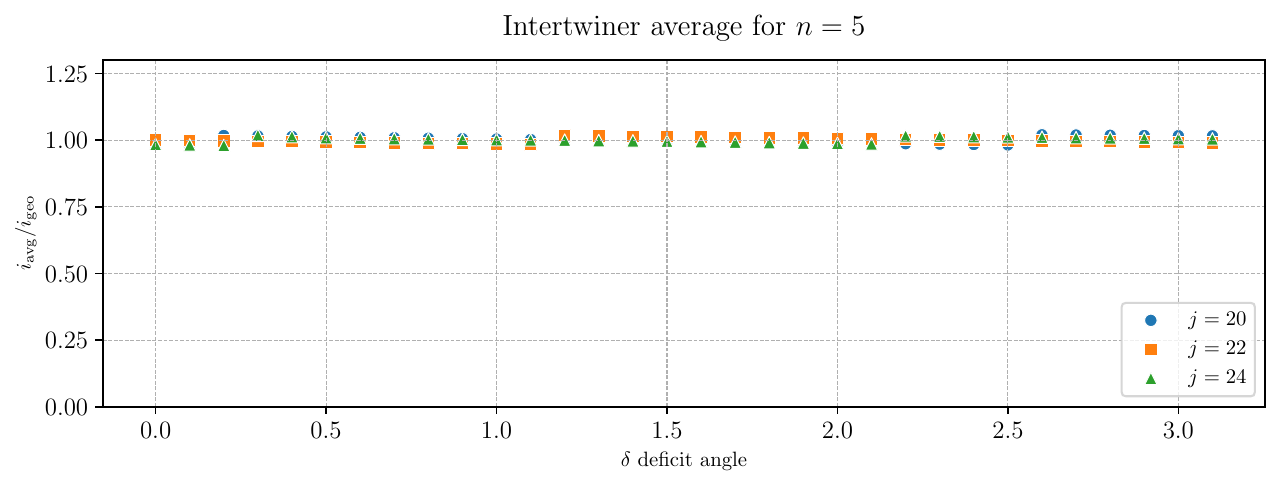}
    \caption{$n=5$}
\end{subfigure}

\par\medskip

\begin{subfigure}[b]{0.48\textwidth}
    \centering
    \includegraphics[width=\textwidth]
    {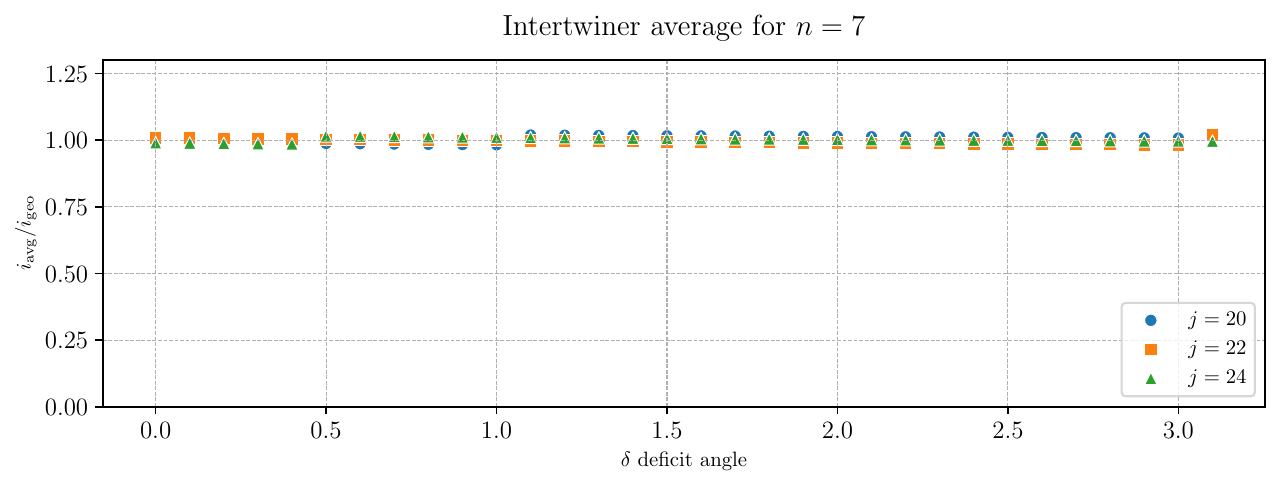}
    \caption{$n=7$}
\end{subfigure}
\hfill
\begin{subfigure}[b]{0.48\textwidth}
    \centering
    \includegraphics[width=\textwidth]
    {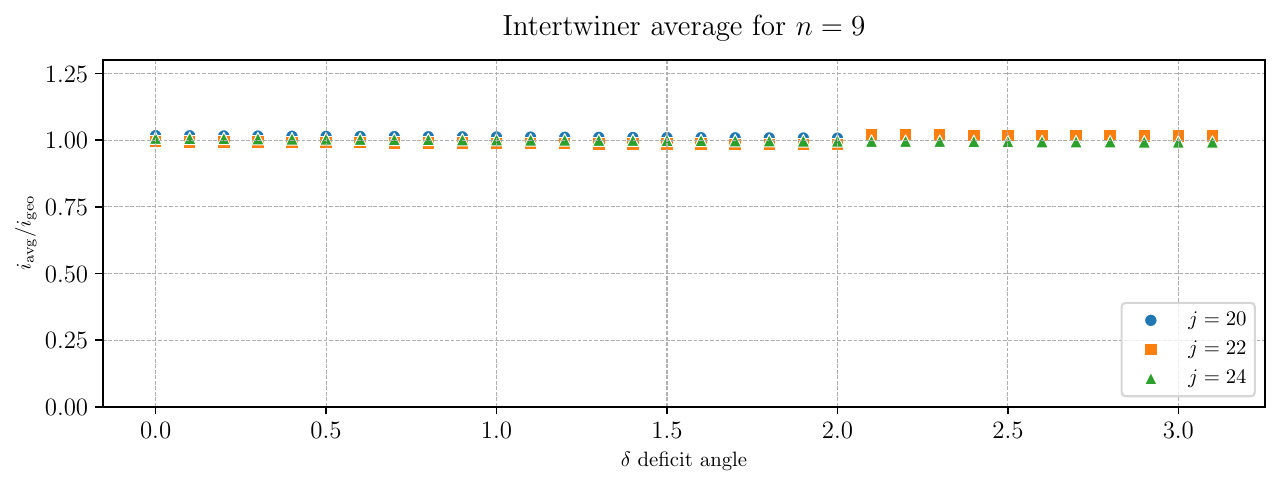}
    \caption{$n=9$}
\end{subfigure}

\caption{
Ratio $\operatorname{Re}\langle i_{\text{b}}\rangle/i_{\mathrm{classical}}$
as a function of the deficit angle $\delta$ for the triangulations
$\Delta_3$, $\Delta_5$, $\Delta_7$, and $\Delta_9$
after application of the finite-window intertwiner projector.
}
\label{fig:intertwiner_average_all_hard}
\end{figure*}

\begin{figure*}[tp]
\centering

\begin{subfigure}[b]{0.48\textwidth}
    \centering
    \includegraphics[width=\textwidth]
    {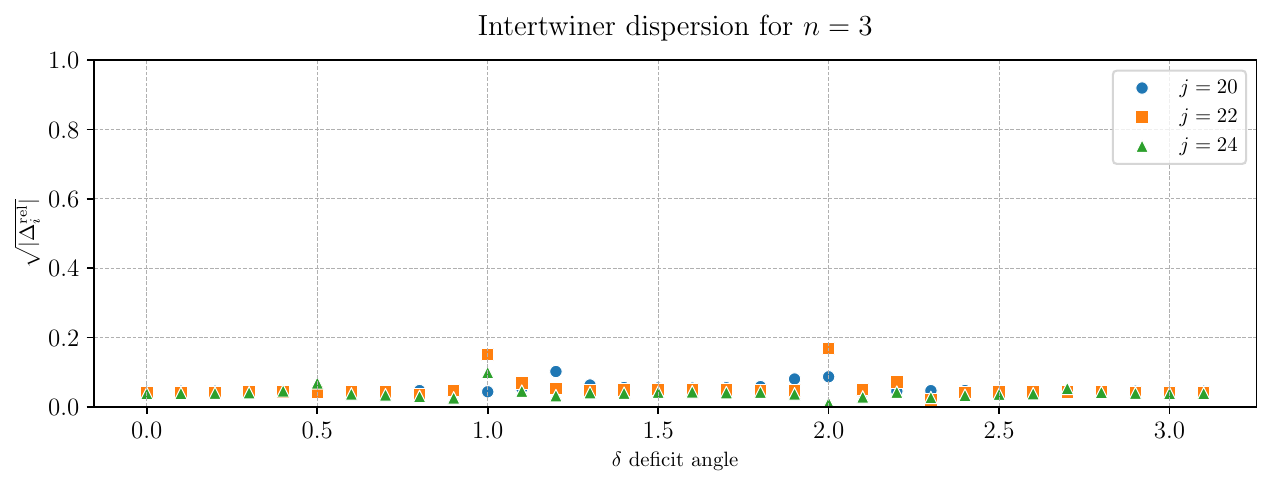}
    \caption{$n=3$}
\end{subfigure}
\hfill
\begin{subfigure}[b]{0.48\textwidth}
    \centering
    \includegraphics[width=\textwidth]
    {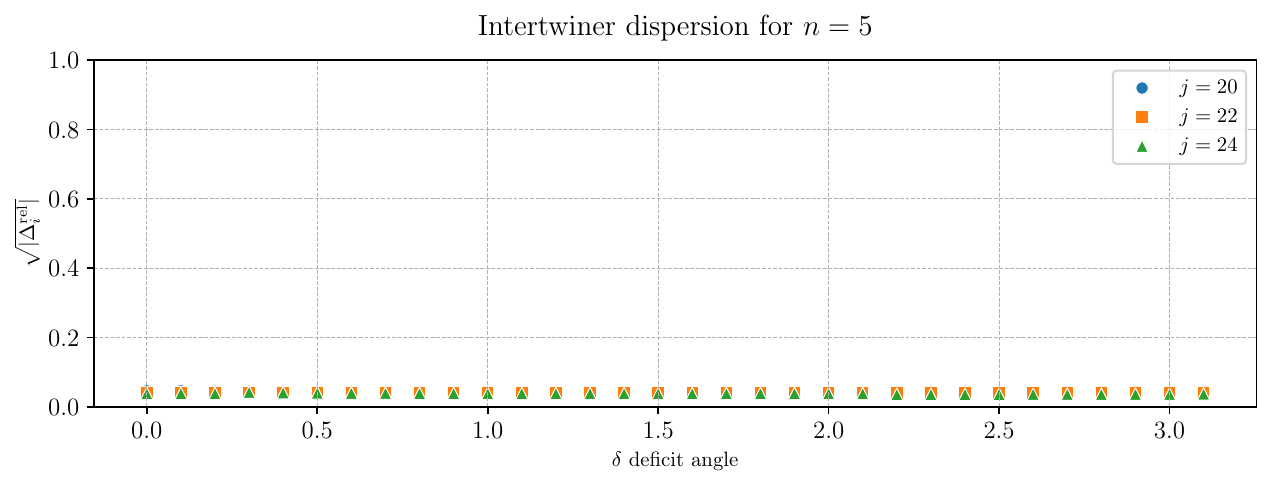}
    \caption{$n=5$}
\end{subfigure}

\par\medskip

\begin{subfigure}[b]{0.48\textwidth}
    \centering
    \includegraphics[width=\textwidth]
    {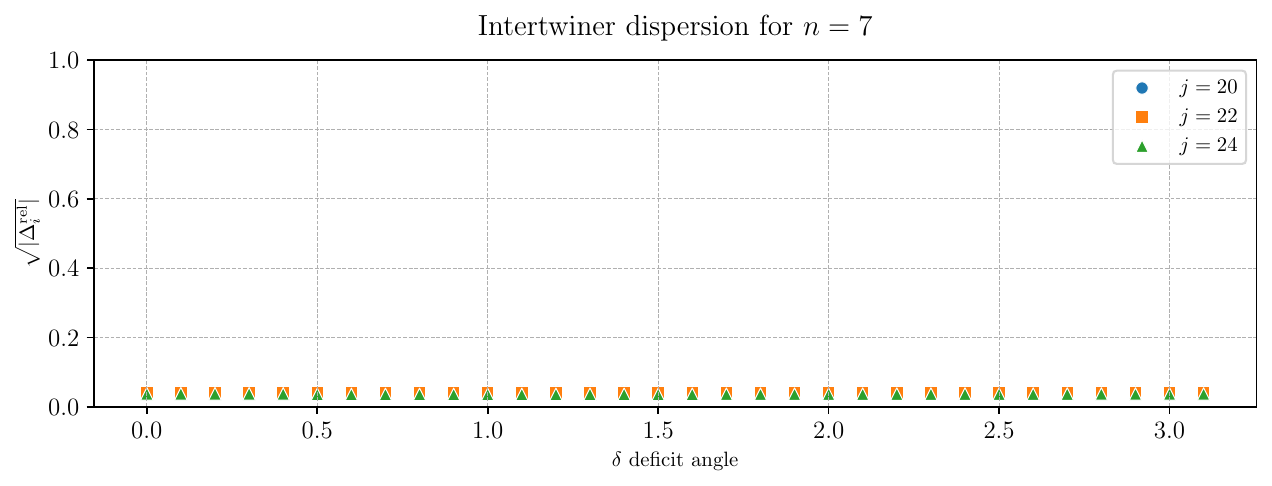}
    \caption{$n=7$}
\end{subfigure}
\hfill
\begin{subfigure}[b]{0.48\textwidth}
    \centering
    \includegraphics[width=\textwidth]
    {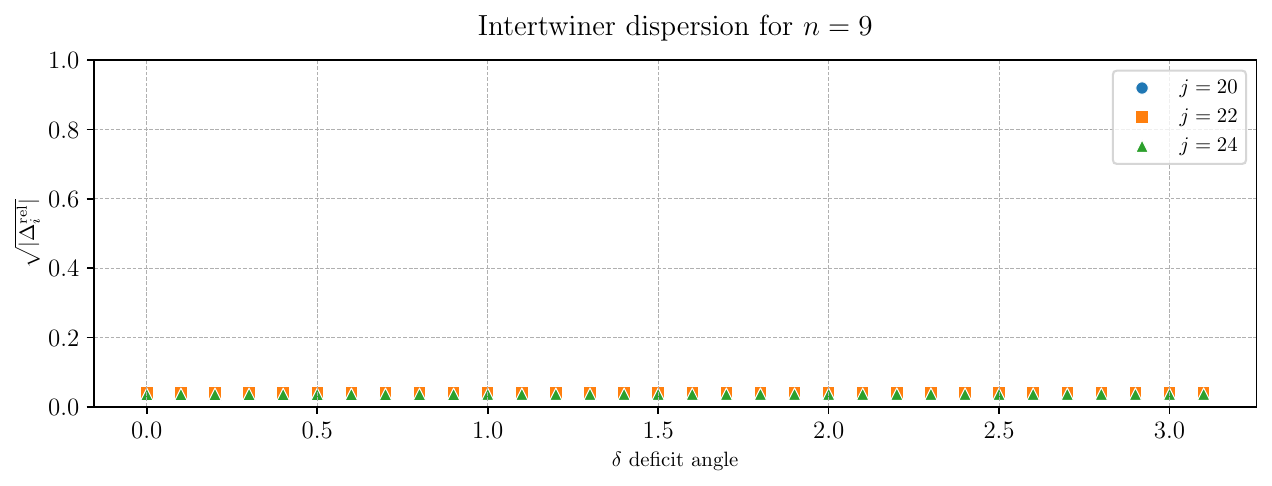}
    \caption{$n=9$}
\end{subfigure}

\caption{
Relative dispersion
$D_i=\sqrt{\left|\langle i_{\text{b}}^2\rangle/\langle i_{\text{b}}\rangle^2-1\right|}$
of the projected intertwiner amplitude as a function of the deficit angle $\delta$
for the triangulations $\Delta_3$, $\Delta_5$, $\Delta_7$, and $\Delta_9$.
}
\label{fig:intertwiner_variance_all_hard}
\end{figure*}

Several observations can be made from the comparison (see \cref{tab:projector_comparison}). First, all projected amplitudes substantially modify the unrestricted result for the $n=5$ triangulation, eliminating the pathological negative value observed in the full amplitude. Second, the resulting expectation values depend nontrivially on the choice of projector. While all tested prescriptions improve the $n=5$ result relative to the unrestricted case, the magnitude of the correction varies significantly. The comparison therefore indicates that the observed behavior depends not only on the presence of an intertwiner restriction, but also on the detailed form of the projector used to implement it.

The finite-window projector serves as the reference prescription for the main analysis. In the cases listed in \cref{tab:projector_comparison}, it retains a ratio near unity for $n=3$ and brings the $n=5$ ratio much closer to the geometric target. The alternative weights give quantitatively different responses, illustrating the role of the restriction prescription. Their geometric interpretation and a wider comparison across parameters are directions for further study.

\subsection{Projected Intertwiner Observables} \label{app:numerics:sec:projected_intertwiner_results}

The projector in \cref{sec:ch5:numerics} restricts one intertwiner channel around the classical reference from \cref{app:geometry}. The plots in \cref{fig:intertwiner_average_all_hard,fig:intertwiner_variance_all_hard} show the real parts of the resulting expectation-value ratios and the relative dispersion $D_i$, allowing a direct comparison of the projected amplitude-weighted moments with their targets.

The plotted ratios in \cref{fig:intertwiner_average_all_hard} remain close to unity for $n=5$, $n=7$, and $n=9$ throughout the range considered. For $n=3$, the largest residual deviations are approximately $10\%-15\%$. 

The relative dispersion $D_i$ of the projected intertwiner amplitude is shown in \cref{fig:intertwiner_variance_all_hard} and are generally small.

The near-classical intertwiner ratios are consistent with the finite window centered on $i_{\text{classical}}$. The complementary bulk-spin response is discussed in \cref{sec:ch5:results:bulk_proj}. Further comparisons of window widths, center offsets, and recoupling channels would help distinguish features of the amplitude response from those tied to the chosen classical input.

\clearpage

\renewcommand{\bibfont}{\small\setlength{\baselineskip}{10.3pt}}
\bibliography{Bibliography}

\end{document}